\documentclass[12pt]{article}
\usepackage{amsmath}
\usepackage{graphicx}
\usepackage{algorithm}
\usepackage{enumitem}
\usepackage{algpseudocode}
\usepackage{multirow}
\usepackage{comment}
\usepackage{booktabs} 
\usepackage{amssymb, amsthm}
\usepackage{dsfont}
\usepackage{xcolor}
\usepackage{subcaption}
\newtheorem{theorem}{Theorem}[section]
\newtheorem{lemma}[theorem]{Lemma}
\newtheorem{corollary}[theorem]{Corollary}


\newtheorem{definition}{Definition}[section]

\newtheorem{prop}{Proposition}[section]
\newtheorem{remark}{Remark}[section]
\usepackage[round]{natbib}  

\newcommand{\blind}{1}

\begin{document}

\def\spacingset#1{\renewcommand{\baselinestretch}%
{#1}\small\normalsize} \spacingset{1}


\if1\blind
{
  \title{\bf Multivariate Discrete Generalized Pareto Distributions: Theory, Simulation, and Applications to Dry spells}
  \author{Samira Aka
    \hspace{.2cm}\\
    IPSL, Univ. Paris-Saclay; ESSEC Business School, CREAR; Square Management\\
    {\tt  
https://orcid.org/0009-0007-9714-8472}\\
    and \\
    Marie Kratz \\
    ESSEC Business School, IDO dpt \& CREAR, Cergy-Pontoise\\
    and \\
    Philippe Naveau \\
   Laboratoire des sciences du climat et de l'environnement, \\ Université Paris-Saclay, CNRS, CEA, UVSQ,
91191 Gif-sur-Yvette, 
France\\
{\tt  
https://orcid.org/0000-0002-7231-6210}
    }
    \date{}
  \maketitle

} \fi

\if1\blind
{
  \bigskip
  \bigskip
} \fi

\bigskip
\begin{abstract}
\noindent 
This article extends the multivariate extreme value theory (MEVT) to discrete settings, focusing on the generalized Pareto distribution (GPD) as a foundational tool. The purpose of the study is to  
enhance the understanding of extreme discrete count data representation, particularly for discrete exceedances over thresholds, defining and using multivariate discrete Pareto distributions (MDGPD). Through theoretical results and illustrative examples, we outline the construction and properties of MDGPDs, providing practical insights into simulation techniques and data fitting approaches using recent likelihood-free inference methods. This framework broadens the toolkit for modeling extreme events, offering robust methodologies for analyzing multivariate discrete data with extreme values. To illustrate its practical relevance, we present an application of this method to drought analysis, addressing a growing concern in Europe.
\end{abstract} 

\noindent%
{\it Keywords:} exceedances over the threshold, drought analysis, neural Bayes estimator
\vfill
\newpage
\spacingset{2} 
\spacingset{1} 

\section{Introduction}
\label{sec:intro}

Extreme value theory (EVT) is a branch of statistics that focuses on understanding and modeling the behavior of extreme events based on the highest (or lowest) values observed in a dataset. The pioneering work on EVT is the Fisher–Tippett–Gnedenko theorem (see \cite{Fisher1928} and \cite{gnedenko1943}), which states that the maximum of a sample of independent, identically distributed continuous random variables, after proper renormalization, converges in distribution towards a three-parameter distribution, the Generalized Extreme Value (GEV) distribution, which corresponds to three types : Fréchet, Weibull and Gumbel. 
Samples with such a property are said to belong to the maximum domain of attraction (MDA) of the given limiting distribution. 
Since then, the theory has been extensively developed to include diverse methods for modeling extreme events, including sample maxima, counts of exceedances over thresholds, and the analysis of tail behavior; see e.g. the books by \cite{Leadbetter1983, Resnick1987, Embrechts1997, Coles2001, Beirlant2004, deHaanFerreira2006, Resnick2007, Falk2019, Carvalho2025}. 
In particular, when analyzing exceedances, defined as the values exceeding a given high threshold, the generalized Pareto distribution (GPD) often serves as an effective approximation. 
The GPD is defined by its cumulative distribution function (cdf):
\begin{equation}
    F(y;\sigma,\xi) =
    \begin{cases}
        1-(1+\xi\frac{y}{\sigma})_+^{-\frac{1}{\xi}}, & \text{for $\xi\neq 0$},\\
        1-e^{-\frac{y}{\sigma}}, & \text{for $\xi=0$},
    \end{cases}
\end{equation}
where $(y)_+ = \max(y,0)$, and $\sigma > 0$, $-\infty < \xi< +\infty$ represent the scale and shape parameters of the distribution, respectively. The parameter \(\xi\) determines the tail behavior, with \(\xi > 0\) indicating heavy tails (Fréchet MDA), $\xi=0$ (Gumbel MDA) and \(\xi < 0\) implying bounded support (Weilbull MDA).
In the univariate case, it has been proven that the distribution of exceedances above a large threshold $u$ for a continuous random variable $Y$ can be approximated as (see \cite{Balkema1974} and \cite{Pickands1975}):
\begin{equation}\label{def:POT}
    \mathbb{P}(Y-u>y|Y\geq u) \approx \Bar{F}(y;\sigma_u,\xi),
\end{equation}
where $\sigma_u$ depends on $u$ and $\Bar{F}=1-F$.
In various domains such as insurance, epidemiology, and natural disaster management, extreme events often take discrete values, such as the number of fires or insurance claims. The Fisher–Tippett–Gnedenko theorem cannot be easily applied to these discrete variables, unless via continuous approximation or by alternative approaches (e.g., via point processes) to model extremes in discrete settings; see e.g. 
\cite{Leadbetter1983} and \cite{Resnick1987,Resnick2007}. Yet, in the last decade, there has been a growing research movement on discrete extreme distributions. In this context, \cite{Chakraborty2015} provided an extensive survey of existing discretizations of continuous extreme or non-extreme distributions. When considering discrete distributions, a significant challenge arises as they do not necessarily possess a defined MDA. \cite{hitz_davis_samorodnitsky_2024} proposed to overcome this limitation by extending the MDA definition. 
This class has the property that, if a continuous random variable $Y$ (with cdf $F$) belongs to a MDA, then the discrete random variable $\lceil Y\rceil$ (using the ceiling function $\lceil.\rceil$) belongs to this larger class.
Within this class, the conditional discrete probability density  function (pdf) is defined,  by
\begin{equation}\label{def:DGPD}
    \mathbb{P}(\lceil Y\rceil-m=n|\lceil Y\rceil\geq m) = \mathbb{P}(Y-m\geq n|Y\geq m) - \mathbb{P}(Y-m\geq n+1|Y\geq m),
\end{equation}
which can be approximated, for $n$ any non-negative integer and $m$ a large integer, by the following discrete pdf defined using F,
\begin{equation}\label{eq:approx_discrete_pdf}
F(n+1;\sigma,\xi) - F(n;\sigma,\xi).
\end{equation}
This so-called discrete GPD (DGPD) has been used to model discrete exceedances over a threshold in some univariate applications. For instance, \cite{Prieto2014} modeled road accident blackspots in Spain and compared it to a classical discrete distribution, the negative binomial, demonstrating its usefulness through classical testing methods. In the same way, \cite{Buddana2014} derived several properties of the DGPD, including stability, infinite divisibility, and the probability generating function, among others, to enhance understanding of this distribution. Similarly, \cite{ahmad2024} incorporated an extended GPD model with zero inflation to account for excessive zero counts, using it within a regression framework for counting avalanches. Lately, \cite{daouia_stup_23} applied the univariate DGPD to study the tail behavior of SARSCoV-2 cluster cases, showing it to be fat-tailed in most countries surveyed. These papers highlight a demand from the applied community to model discrete exceedances. So far, research appears to have focused solely on univariate cases.

Although the univariate EVT framework provides valuable insight into the tail behavior of marginals distributions, most real-world phenomena involve multiple dependent variables that require an extension to multivariate extremes to capture their simultaneous behavior. 

In the continuous setting, multivariate extreme value theory (MEVT) originates from the study of the limiting joint distribution of normalized componentwise maxima of random vectors. This theory has evolved into a comprehensive framework for modeling and understanding the joint behavior of extreme events across multiple variables (see, e.g., the books by \cite{Coles2001, Beirlant2004, deHaanFerreira2006, Resnick2007, Falk2019, Carvalho2025}). More recently, the development of the continuous multivariate generalized Pareto distribution (GPD) has extended MEVT beyond block maxima, enabling the modeling of threshold exceedances in a multivariate context. This has proven especially useful in various applications, such as climatology, also of interest here (see, e.g., \cite{Kiriliouk:Naveau:2020}).

 Following the seminal paper that defined the multivariate generalized Pareto distribution (MGPD), namely \cite{rootzen_tajvidi2006}, several subsequent papers provided properties for fitting this particular distribution, such as \cite{rootzen_segers_wadsworth,rootzenSegerswadsworth_2} and \cite{Kiriliouk2019}. 
 However, to our knowledge, this distribution has not yet been extended to the study of a discrete multivariate setting. This is the main objective of this paper, motivated by an application in climatology: modeling dry spells - sequences of consecutive days with little or no precipitation - a problem that is discrete by nature.
 
\begin{figure}[H]
    \centering
    \includegraphics[width=0.9\textwidth]{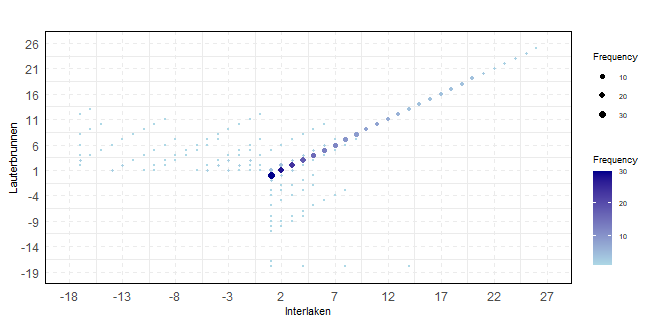} 
    \caption{\sf \small Scatter plot of dry spells exceeding the 99\% quantile for two Swiss stations 7 km apart (which correspond respectively to 17 days for Interlaken and 18 days for Lauterbrunnen). The bigger and darker the points, the longer the dry spells.}
    \label{fig:DS_close_intro}
\end{figure}

As highlighted in IPCC reports \citep[see, e.g.][]{ipcc22}, dry spells are key to understanding past, present, and future droughts with important consequences in water availability and fire risks. 
 Since the lengths of these spells are countable, they must be represented by a discrete distribution. This is illustrated in Figure \ref{fig:DS_close_intro}, which displays a scatter plot of joint extreme dry spells based on precipitation data recorded in Switzerland from 1930 to 2014 used by \cite{Pasche_2022}. The data correspond to two nearby locations:  
the stations of Interlaken (x-axis) and Lauterbrunnen (y-axis), 7 km away from each other.
 Here, dry spells represent numbers of consecutive days without precipitation, and the term {\em joint extremes} means that either Interlaken or Lauterbrunnen has recorded a dry spell length greater than their 99\% quantiles. More precisely, the origin $(0,0)$ denotes a joint dry spell of length 17 days in Interlaken and 18 days in Lauterbrunnen. The blue points correspond to the frequency with their size and shade increasing with the frequency. A large  dark blue point indicates high frequency, while a small light blue point a low frequency. This figure highlights the need for a discrete bivariate distribution to accurately capture the observed joint behavior, raising the key question of which distribution best represents this dependence. It also motivates the theoretical developments and inference approaches presented in the subsequent sections. 

 In Figure~\ref{fig:DS_close_intro}, one can observe that the joint frequency is higher along the diagonal indicating that  extreme dry spells can jointly occur at these two nearby stations. 
\vspace{1ex}
The paper is organized as follows: The end of Section~\ref{sec:intro} recalls some notions of MGPD (continuous setting). Section~\ref{sec:MGDPD} focuses on the definition of the MDGPD. Section~\ref{sec:simulations_of_MDGPDs} provides the key steps to simulate it. Section~\ref{sec:inference} presents inference methods applied first to simulated data, then to the dry spells data displayed in Figure~\ref{fig:DS_close_intro}. Proofs of all results are developed in the appendix.\\[1ex]
Before concluding this introduction, let us recall some notation, a mathematical definition of the MGPD, and a new proposition to define a non-standard MGPD. 
Bold face symbols denote vectors in $\mathbb{K}^d$ where $\mathbb{K}=\mathbb{R},\mathbb{Z}$ or $\mathbb{N}$, for example, $\boldsymbol{0} = (0, \ldots, 0)$ and $\boldsymbol{1} = (1, \ldots, 1)$. Mathematical operations such as addition, multiplication, and exponentiation are to be interpreted componentwise when applied to vectors. For instance, for $\boldsymbol{x}, \boldsymbol{\xi} \in \mathbb{R}^d$, we write $(\boldsymbol{1 + \xi x})^{\boldsymbol{-\frac{1}{\xi}}}$ for the vector 
$\left((1 + \xi_1 x_1)^{-\frac{1}{\xi_1}}, \ldots, (1 + 
\xi_d x_d)^{-\frac{1}{\xi_d}}\right).$
 We let $a \wedge b$ denote $\min(a, b)$, whereas for vectors, the minimum and the maximum are taken componentwise. We write $\mathcal{L}(\boldsymbol{N})$ for the law of a random  vector $\boldsymbol{N}$.
 Writing $\left(\boldsymbol{Z}\nleqslant \boldsymbol{u}\right)$ means that at least one coordinate of $\boldsymbol{Z}$ exceeds the corresponding coordinate of $\boldsymbol{u}$, i.e., $\bigvee_{j=1}^d \left( Z_j > u_j \right)
$. $\mathds{1}_{A}$ is the indicator function, equal to 1 if A is true and to 0 otherwise. 

 Concerning the continuous MGPD, various definitions exist ; see, e.g., the survey by Naveau and Segers, Chapter 5 in \cite{Carvalho2025}. For this work, we opt for the following one, by \cite{rootzen_segers_wadsworth}. 
 
\begin{definition}\label{def: st MGPD}[\cite{rootzen_segers_wadsworth}, Theorem 7] \label{th_Z+E_Rootzen_et_al.}
A random vector $\boldsymbol{Z}\in  \mathbb{R}^d$ is said to follow a standard multivariate generalized Pareto distribution $MGPD(\boldsymbol{1},\boldsymbol{0},\boldsymbol{S})$ if
\begin{itemize}
    \item The random variable $E:=\max(\boldsymbol{Z})$ follows a unit exponential distribution, i.e. \\$\mathbb{P}(E> x)=e^{-x}, x\geq0;$
    \item The random vector $\boldsymbol{S}=\boldsymbol{Z}-E,$ named spectral random vector, is independent of $E$ and satisfies 
   $$
\mathbb{P}\left(\underset{1\leq i\leq d}{\max}{S}_i=0\right) = 1 \quad \text{and} \quad \mathbb{P}({S}_j > \infty) > 0,\ \forall j = 1, \ldots, d.
$$
\end{itemize}
The random vector $\boldsymbol{X}$ defined by
\begin{equation*}
    \boldsymbol{X}=\boldsymbol{\sigma}\frac{e^{\boldsymbol{\xi Z}}-1}{\boldsymbol{\xi}}, \text{with $\boldsymbol{\sigma}>\boldsymbol{0}$ and $\boldsymbol{\xi}\in \mathbb{R}^d$,}
\end{equation*}
is called   a non-standard $MGPD(\boldsymbol{\sigma},\boldsymbol{\xi},\boldsymbol{S})$.
\end{definition}

In the definition of $\boldsymbol{S}$, we impose $\mathbb{P}(S_j = -\infty) = 0$ for all $j$, in contrast to the more general formulation recalled by Naveau and Segers in~\cite{Carvalho2025}, where $\boldsymbol{Z} \in [-\infty, \infty)^d$. This assumption is made to avoid complications arising from certain independent cases and to simplify the analysis.\\  
The cdf of $\boldsymbol{Z}$ is given by
\begin{equation}\label{eq:cdf_MGPD}
H(\boldsymbol{z})=1-\mathbb{E}\left(1\wedge  e^{\max(\mathbf{S}-\mathbf{z})}\right).   
\end{equation}
A key feature of the MGPD is that this distribution is threshold invariant in the sense that $[\boldsymbol{Z}|\boldsymbol{Z}\nleqslant\boldsymbol{u}]$ is still a MGPD.

Moreover, we show in Proposition \ref{prop:Z_over_Gamma}
that the transformation from a standard MGDP to a non-standard one can also be obtained using an independent random Gamma variable. This  first result, which proof is given in Appendix \ref{App:prop1}, will help us move from the continuous framework to the discrete one, as developed in Section \ref{sec:MGDPD}.
\begin{prop}\label{prop:Z_over_Gamma}
Let $\Lambda$ be  a $Gamma(\alpha, \beta)$ distributed random variable  and  $\boldsymbol{Z}$   a standard $MGPD(\boldsymbol{1},\boldsymbol{0},\boldsymbol{S})$     independent of $\Lambda$. 
Then, the ratio  $\boldsymbol{Z} /\Lambda$ follows a $MGPD\left(\boldsymbol{\frac{\beta}{\alpha}},\boldsymbol{\frac{1}{\alpha}},\boldsymbol{S}\right).$
\end{prop}

\section{Multivariate discrete generalized Pareto distributions}
\label{sec:MGDPD}


To define a standard discrete MGPD, a first step is to study the integer part of $\boldsymbol{Z},$ a standard MGPD$(\boldsymbol{1},\boldsymbol{0},\boldsymbol{S})$.
By construction, $\max(\boldsymbol{Z})=E,$ and this property remains true for the integer part of $\boldsymbol{Z}$:
\begin{equation}\label{eq:max_Z}
\max(\lceil\boldsymbol{Z}\rceil) = \lceil E\rceil 
\end{equation} (see Appendix \ref{app:sec2}). It is also known from \cite{Steutel1989} that $\lceil E\rceil$ follows a geometric distribution with parameter $(1 - e^{-1})$, and that the integer and fractional parts of an exponential variable are independent, so $E$ is independent of $\boldsymbol{S}$. We can then deduce that $\lceil E\rceil$ is independent of $\lceil\boldsymbol{S} - {E}\rceil$ and introduce the following definition.
\begin{theorem}\label{th:MDGPD}
A random vector $\boldsymbol{N}\in  \mathbb{Z}^d$ is said to follow a standard multivariate discrete generalized Pareto distribution $MDGPD(\boldsymbol{1},\boldsymbol{0},\boldsymbol{S})$ if 
\begin{itemize}
    \item The discrete random variable $G:=\max(\boldsymbol{N})$ follows a geometric distribution with parameter $1-e^{-1}$
    \begin{equation*}
        \mathbb{P}(G\leq n)=1-e^{-n}, \quad n\in \mathbb{N}^*;
    \end{equation*}
    \item  The random vector $\boldsymbol{S}=\boldsymbol{N}-G$ is independent of $G$ and satisfies      $$\mathbb{P}\left(\underset{1\leq i\leq d}{\max}{S}_i=0\right)=1 \quad \text{and} \quad \mathbb{P}\left(\underset{1\leq i\leq d}{\min}{S}_i>-\infty\right)=1.$$
\end{itemize}
\end{theorem}
We call $\boldsymbol{S}=\boldsymbol{N}-G$  a \textit{discrete spectral random vector}
 and we observe that  the cdf of $\boldsymbol{N}$ is given, for $\boldsymbol{n}\in \mathbb{Z}^d$, by
\begin{equation*}\label{eq:cdf_MDGPD}
1-\mathbb{E}(1\wedge e^{\max(\mathbf{S}-\mathbf{n})}).  
\end{equation*}
\noindent Note that $\boldsymbol{N}$ can also be written as 
\begin{equation}\label{eq:T-max(T)}
\boldsymbol{N}\overset{d}{=}\boldsymbol{T} - \max(\boldsymbol{T}) + G,
\end{equation}
where $\boldsymbol{T},$ called a generator, is any  discrete $d$-dimensional random vector.  To understand this decomposition \eqref{eq:T-max(T)} of $\boldsymbol{N}$ into three components, let us focus on the bivariate case with $\boldsymbol{N}=(N_1,N_2)$ and $\boldsymbol{T}=(T_1,T_2)$.  

As  $N_1-N_2=T_1-T_2$, introducing the notation  $\Delta=T_1-T_2,$ we can write
\begin{equation*}\label{eq:MDGPD_Delta}
    \begin{cases}
        N_1 = G + \Delta\mathds{1}_{\left(\Delta<0\right)}, \\
        N_2 = G - \Delta\mathds{1}_{\left(\Delta\geq 0\right)}.
    \end{cases}
\end{equation*}
From this, we can derive properties of the marginals (see Appendix~\ref{app:sec2} for the proof).
\begin{corollary}{\textbf{Marginals properties}}\label{cor:N_marg}
\\For any $n\in \mathbb{Z}$,
\begin{equation*}
     \mathbb{P}\left(N_1=n\right)=\left(1-e^{-1}\right)e^{-n+1}\left(\mathbb{E}\left[e^{\Delta}\mathds{1}_{(\Delta<\min(0,n))}\right]+\mathbb{P}\left(\Delta\geq 0\right)\mathds{1}_{(n>0)}\right),
\end{equation*}
and 

\begin{equation}\label{eq: P(N1>n)}
    \mathbb{P}\left(N_1> n\right)= \mathbb{P}\left(n\leq\Delta<0\right)\mathds{1}_{(n<0)}+e^{-n}\mathbb{E}\left( e^{\Delta}\mathds{1}_{(\Delta\leq n-1)}\right)\mathds{1}_{(n\leq0)}+\mathbb{P}\left(\Delta\geq 0\right)\left(\mathds{1}_{(n\leq0)}+e^{-n}\mathds{1}_{(n\geq 1)}\right).
\end{equation}

This gives a conditional marginal invariance (memoryless property of the geometric distribution) with 
$$
\mathbb{P}\left(N_1> n|N_1>0\right)= e^{-n}=\mathbb{P}\left(G> n\right).
$$ 
Since $N_2$ is constructed symmetrically to $N_1$ with respect to $\Delta$, its distribution is derived accordingly. 

More generally, in dimension $d>2$, for $\boldsymbol{N}=(N_1,\ldots,N_d)$ and $\boldsymbol{T}=(T_1,\ldots,T_d),$ introducing \[\Delta_i = T_i - \underset{j \neq i}{\max}{}\, T_j \quad\text{for} \quad i=1,\ldots,d,\]
we obtain from Equation~\eqref{eq:T-max(T)} the following definition:
\begin{equation*}
    N_i=G+\Delta_i\mathds{1}_{(\Delta_i<0)}, \text{ for $i=1,\ldots,d$},
\end{equation*}
and, similarly as in dimension $2$, we have, for all $i=1,\ldots,d$, 
$$
\mathcal{L}(N_i|N_i>0)=\mathcal{L}(G).
$$ 
\end{corollary}
As for its continuous counterpart, the stability of the marginal threshold can be generalized for the MDGPD in the following way. 
 \begin{corollary}\label{prop:Dist_AZ}
  Let $\boldsymbol{N}$ be a MDGPD$(\boldsymbol{1},\boldsymbol{0},\boldsymbol{S})$ and let $\boldsymbol{A}=(a_{ij})$ be a non-negative integer-valued matrix $\in \mathbb{N}^{n\times d}$ such that $\mathbb{P}(\sum_{j=1}^d a_{ij}N_j>0)>0,\forall i=1,\ldots,n.$ Let $\boldsymbol{m}\in \mathbb{N}^n.$
  Then, $$
  \mathcal{L}(\boldsymbol{AN-m}|\boldsymbol{AN} \nleqslant \boldsymbol{m})=MDGPD(\mathbf{A1}, \mathbf{0}, \boldsymbol{S_n}),
  $$
which is defined in $\mathbb{N}^n$ by
\begin{equation}\label{eq:linear_comb}
     1-\frac{\mathbb{E}\left[1\wedge \exp\left\{\left\lceil\underset{1\leq i \leq n}{\max}\left(U_i-\frac{k_i}{\sum_{j=1}^d a_{ij}}\right)\right\rceil\right\}\right]}{\mathbb{E}\left[1\wedge \exp\left\{\left\lceil\underset{1\leq i \leq n}{\max}U_i\right\rceil\right\}\right]},
\end{equation}
where the distribution of $\boldsymbol{S_n}$ is
  \begin{equation}\label{eq:def_S_n}
      \mathbb{P}(\boldsymbol{S_n}= \boldsymbol{l})=\frac{\mathbb{E}\left[\exp\left\{\left\lceil\underset{1\leq i \leq n}{\max}U_i\right\rceil\right\}\mathds{1}_{\left(\lceil\boldsymbol{U}-\max(\boldsymbol{U})\rceil=l\right)}
\right]}{\mathbb{E}\left[\exp\left\{\left\lceil\underset{1\leq i \leq n}{\max}U_i\right\rceil\right\}\right]},
  \end{equation}
and $U_i = \frac{\sum_{j=1}^d a_{ij} S_j -  m_i}{\sum_{j=1}^d a_{ij}}, \ \forall \ i = 1, \ldots, n$.
\end{corollary}
Taking the identity matrix for $A$ with $\boldsymbol{m}\in\mathbb{N}^n$ fixed, gives the threshold stability 
    \begin{equation*}
        \mathcal{L}(\boldsymbol{N-m}|\boldsymbol{N} \nleqslant \boldsymbol{m})=MDGPD(\mathbf{1}, \mathbf{0}, \boldsymbol{S_n}),
    \end{equation*}
and the corresponding cdf is given, for $\boldsymbol{k}\in \mathbb{Z}^n$, by
\begin{equation*}\label{eq:threshold_stability_standard}
H(\boldsymbol{k-m}) = 1 - \frac{\mathbb{E}\left(1 \wedge e^{\max(\boldsymbol{S} - \boldsymbol{k} )}\right)}{\mathbb{E}\left(1 \wedge e^{\max(\boldsymbol{S} -\boldsymbol{m})}\right)}.
\end{equation*}
With these fundamental concepts in place for the standard MDGPD $\boldsymbol{N}$, we now move on to the non-standard case. 
In the continuous case, this step was implemented by using the non-linear  transformation $    \boldsymbol{X}=\boldsymbol{\sigma}\frac{e^{\boldsymbol{\xi Z}}-1}{\boldsymbol{\xi}}
$ (see Definition \ref{def: st MGPD}). Such a trick does not work here, as the standard $\boldsymbol{N}$ takes integer values, but 
$\displaystyle \boldsymbol{\sigma}\frac{e^{\boldsymbol{\xi N}}-1}{\boldsymbol{\xi}}$ 
does not if one component of $\boldsymbol{\xi}$ is non-null.
Therefore, we first define non-standard discrete MGPDs by imposing their cdf as done in Definition \ref{def:non-standard_MDGPD}. Then, we show how to construct random vectors having this type of distribution and describe some of their properties.  
\begin{definition}\label{def:non-standard_MDGPD}
A   random vector $\boldsymbol{M}\in  \mathbb{Z}^d$ is said to follow a non-standard 
$MDGPD\left(\boldsymbol{\sigma}, \boldsymbol{\xi}, \boldsymbol{S}\right)$ 
if, for any integer valued  vector $\boldsymbol{k}$, 
\begin{align*} 
\mathbb{P}\left(\boldsymbol{M} \leq \boldsymbol{k}\right) 
= 1 - \mathbb{E} \left( 
1 \wedge e^{\max \left( \boldsymbol{S} - \boldsymbol{\frac{1}{\xi}} 
\log \left( \boldsymbol{1}  +\frac{\boldsymbol{\xi}\boldsymbol{k}}{\boldsymbol{\sigma}} \right) \right)} \right).
\end{align*}    
\end{definition}
For the special case $\boldsymbol{\xi}=\boldsymbol{0}$ and $\boldsymbol{\sigma}=\boldsymbol{1}$, we have 
$\displaystyle \mathbb{P}\left(\boldsymbol{M} \leq \boldsymbol{k}\right) = 
1 - \mathbb{E} \left(1 \wedge e^{\max \left( \boldsymbol{S} -  \boldsymbol{k}\right) } \right)$.

Building on Theorem~\ref{th:MDGPD} and Corollary~\ref{cor:N_marg}, we now present a generalization of the MDGPD that extends its applicability to linear transformations and threshold shifts.
\begin{prop}{\textbf{Generalization of the MDGPD }}\label{prop:discrete_generalised_gpd}

  Let $\boldsymbol{M}=\frac{\boldsymbol{Z}}{\Lambda},$ where $\boldsymbol{Z}$ and $\Lambda$ are defined as in Proposition \ref{prop:Z_over_Gamma}.
  Let 
  $\boldsymbol{A}=(a_{ij})$ be a matrix $\in \mathbb{N}^{n\times d}$ such that $\mathbb{P}\left(\sum_{j=1}^d a_{ij}N_j>0\right)>0,\forall i=1,\ldots,n,$ and let $\boldsymbol{m}\in \mathbb{N}^n.$
  Then, 
  $$
  \mathcal{L}\left(\left\lceil\boldsymbol{AM}\right\rceil-\boldsymbol{m}|\lceil\boldsymbol{AM}\rceil \nleqslant \boldsymbol{m}\right)=MDGPD\left(\boldsymbol{\frac{\beta}{\alpha}A1}, \boldsymbol{\frac{1}{\alpha}}, \boldsymbol{S_n}\right),
  $$
  with cdf given, for $\boldsymbol{k}\in\mathbb{Z}^d,$ by
\begin{equation}\label{eq:floor(AM)}
1 - \frac{\mathbb{E}\left[1 \wedge \exp\left\{\underset{1\leq i \leq m}{\max} \left(V_i - \alpha \log\left(1 + \frac{k_i}{\beta \sum_{j=1}^d a_{ij}+m_i}\right)\right)\right\}\right]}{\mathbb{E}\left[1 \wedge \exp\left(\underset{1\leq i \leq m}{\max} V_i \right)\right]},
\end{equation}
where $\displaystyle V_i = \frac{\sum_{j=1}^d a_{ij}S_j}{\sum_{j=1}^d a_{ij}} - \alpha \log\left(1 + \frac{m_i}{\beta\sum_{j=1}^d a_{ij}}\right), \forall i = 1, \ldots, n$, 
and \\the distribution of $\boldsymbol{S_n}$ is
\begin{equation}\label{eq:def_S_n_floor(AM)}
\mathbb{P}(\boldsymbol{S_n} \in ) = \frac{\mathbb{E}\left[\exp\left(\underset{1\leq i \leq m}{\max} V_i \right) \mathds{1}_{\{\boldsymbol{V} - \max(\boldsymbol{V}) \in \}} \right]}{\mathbb{E}\left[\exp\left(\underset{1\leq i \leq m}{\max} V_i \right)\right]}.
\end{equation}
\end{prop}
When $\boldsymbol{A}$ is the identity matrix and $\boldsymbol{m}=\boldsymbol{0}$, 
\begin{equation*}
 \left\lceil \frac{\boldsymbol{Z}}{\Lambda} \right\rceil\text{ follows a }MDGPD\left(\boldsymbol{\frac{\beta}{\alpha}}, \boldsymbol{\frac{1}{\alpha}}, \boldsymbol{S}\right).
 \end{equation*}  

 \begin{remark}
Notice that, when $n=1$, 
\[  \mathcal{L}(\left\lceil\boldsymbol{AM}\right\rceil-\boldsymbol{m}|\lceil\boldsymbol{AM}\rceil \nleqslant \boldsymbol{m})\text{ is a univariate discrete GPD} \left(\frac{\beta\sum_{j=1}^d a_j+m}{\alpha},\frac{1}{\alpha}\right)
\]
since we can write
\begin{equation*}
    \mathbb{P}\left(\left\lceil\boldsymbol{AM}\right\rceil-m\leq l|\lceil\boldsymbol{AM}\rceil>m\right)=1-\left(1+\frac{l}{\beta\sum_{j=1}^d a_j+m}\right)^{-\alpha}.
\end{equation*}
 \end{remark} 

It is natural to question whether the use of a discrete distribution is truly necessary, given the availability of continuous models which could be used as approximation, taking their integer part.  
The following proposition addresses this point theoretically, while Section~\ref{sec:inference} and Table~\ref{tab:results_sim} present numerical experiments demonstrating that, although the continuous approximation of discrete data is asymptotically valid (as established in Theorem~\ref{th:DS_over_cont}), it may lead to less accurate estimations when applied to finite sample sizes.
\begin{theorem}
\label{th:DS_over_cont}
 For $\boldsymbol{\sigma}>\boldsymbol{0},\,\boldsymbol{\xi}\geq \boldsymbol{0},$ the probability density and mass functions of the MGPD and MDGPD are asymptotically equivalent as $\boldsymbol{\sigma}$ tends to
infinity. It holds that:
    \begin{equation*}
\lim_{\boldsymbol{\sigma}\rightarrow \boldsymbol{\infty}}\sup_{\boldsymbol{k}\in\boldsymbol{\mathbb{N}^*}}\left|\frac{\mathbb{E}\left[1\wedge e^{\max\left(\boldsymbol{S-\frac{1}{\xi}log(\frac{\xi k}{\sigma}+1)}\right)}\right]}{\mathbb{E}\left[1\wedge e^{\max\left(\boldsymbol{S_D-\frac{1}{\xi}log(\frac{\xi k}{\sigma}+1)}\right)}\right]}-1\right|=0,
    \end{equation*} 
where $\boldsymbol{S}$ and $\boldsymbol{S_D}$ denote the random spectral vector in the continuous and discrete settings, respectively.
\end{theorem}
Theorem~\ref{th:DS_over_cont} suggests that modeling a sample from $\left[\boldsymbol{X}-\boldsymbol{u}|\boldsymbol{X}\nleqslant \boldsymbol{u}\right]$ using a MDGPD or the integer part of a MGPD should not differ too much as long as the estimated scale parameter $\boldsymbol{\hat\sigma}$ is sufficiently large. In practice, when the sample size and threshold grow, the estimated scale parameter might not be large enough for the two distributions to be equivalent. This justifies the need for a different distribution for discrete values.

\section{Simulations of discrete bivariate GPD}
\label{sec:simulations_of_MDGPDs}

From the definitions given in Section~\ref{sec:MGDPD}, we observe a wide range of possibilities for choosing a generator $\boldsymbol{T}$ to simulate an MDGPD. This results in a broad spectrum of distribution families for data modeling. However, such flexibility can make the calibration process particularly challenging. To address this, at least partly, we will rely on likelihood-free models, as discussed in Section~\ref{sec:inference}.

Using the distribution equality \eqref{eq:T-max(T)}, if the vector $ \boldsymbol{T} $ can be efficiently simulated, then a standard $ MGPD $ sample can be generated by adding a geometric random variable $ G $ to $\boldsymbol{T} - \max(\boldsymbol{T})$.
In this section, we focus on the bivariate case for illustration and to simplify the notations. 
Some extensions to the multivariate setup are described in Appendix~\ref{app:dim_3}.

In Equation~\eqref{eq: P(N1>n)}, the difference $\Delta=T_1-T_2$ represents the only random variable needed from $\boldsymbol{T}=(T_1,T_2)$.    
This leads to the following simulation algorithm of discrete bivariate GPD.
\begin{algorithm}
\caption{Simulation of standard bivariate discrete GPD}
\label{algo:known_increments}
\begin{algorithmic}[1]
\Ensure  A sample of   $n$ bivariate standard DGPD  as in Eq. \eqref{eq:T-max(T)}
\Require  Choose a  distribution for $\Delta$

\State Sample  $n$ realizations of   $\Delta$:  $(\Delta_1, \dots , \Delta_n)$
\State Sample  $n$ realizations of  geometrically  distributed  random variables with parameter $1-e^{-1}$ and independently of  $\Delta_i$: $(G_1, \dots, G_n)$
\State Return a bivariate sample of $\boldsymbol{N}_k$ for $k=1,\dots,n$ with, 
\begin{equation*} 
    \begin{cases}
        N_{1,k} = G_k + \Delta_k\mathds{1}_{\left(\Delta_k<0\right)}, \\
        N_{2,k}= G_k - \Delta_k\mathds{1}_{\left(\Delta_k\geq 0\right)}.
    \end{cases}
\end{equation*}
\end{algorithmic}
\end{algorithm} 
The choice of the distribution of $\Delta$ is a key component in the proposed algorithm. Depending on the application context, this distribution can either be specified parametrically or learned directly from the underlying generator \(\boldsymbol{T} = (T_1, T_2)\). Here, we explore three types of Poisson generators \(\boldsymbol{T}\), summarized in Table~\ref{tab:Num_examples}. The choice of the Poisson distribution is motivated by the dry spell dataset under study. Indeed, although we simulated several other discrete distributions for \(\boldsymbol{T}\) within the MDGPD framework, they did not adequately capture the characteristics of the observed real data.
These three examples illustrate how the structure of \(\boldsymbol{T}\) influences the resulting distribution of the discrete MDGPD vector \(\boldsymbol{N} = (N_1, N_2)\), as defined via the representation in \eqref{eq:T-max(T)}.

\begin{figure}[H]
    \centering
        Type (a)\\
\fbox{\begin{subfigure}[b]{0.5\textwidth}
        \centering
        \includegraphics[width=\textwidth]{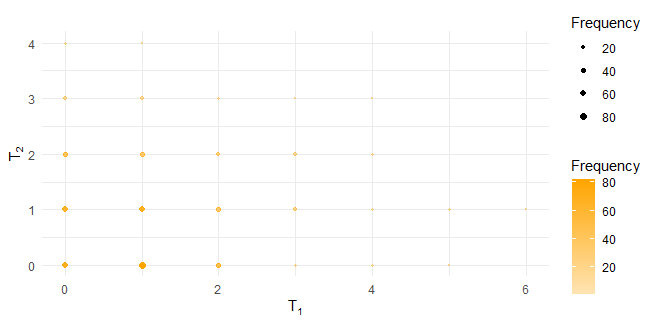}
        \label{fig:T_indep}
    \end{subfigure}
    \hfill
    \begin{subfigure}[b]{0.5\textwidth}
        \centering
        \includegraphics[width=\textwidth]{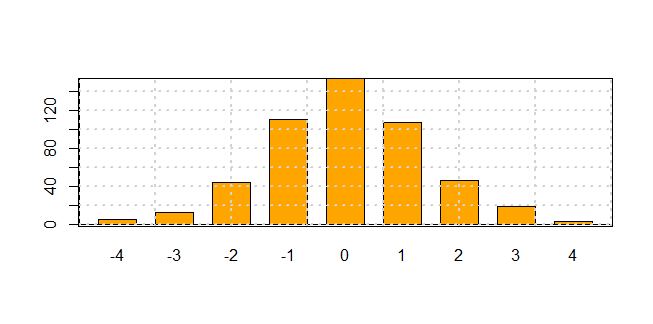}
        \label{fig:delta_indep}
    \end{subfigure}
    }\\
        Type (b)\\
\fbox{
    \begin{subfigure}[b]{0.49\textwidth}
        \centering
        \includegraphics[width=\textwidth]{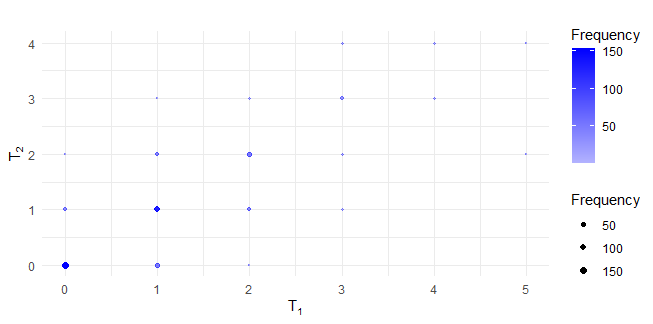}
        \label{fig:T_dep}
    \end{subfigure}
    \hfill
    \begin{subfigure}[b]{0.5\textwidth}
        \centering
        \includegraphics[width=\textwidth]{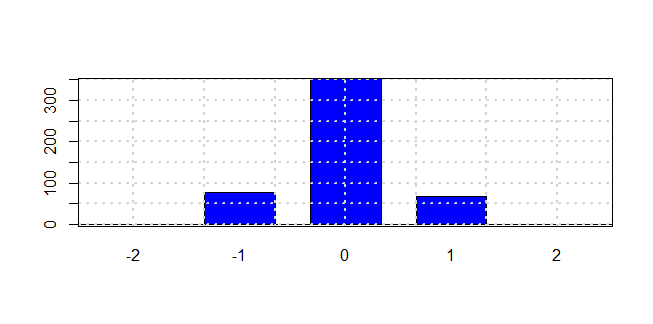}
        \label{fig:delta_dep}
    \end{subfigure}
    }\\
            Type (c)\\
\fbox{
\begin{subfigure}[b]{0.5\textwidth}
        \centering
        \includegraphics[width=\textwidth]{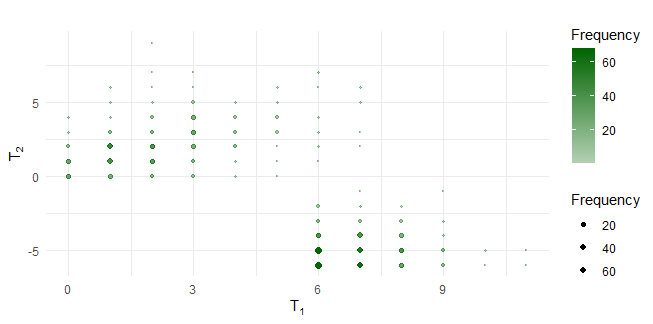}
        \label{fig:T_bimodal}
    \end{subfigure}
    \hfill
    \begin{subfigure}[b]{0.49\textwidth}
        \centering
        \includegraphics[width=\textwidth]{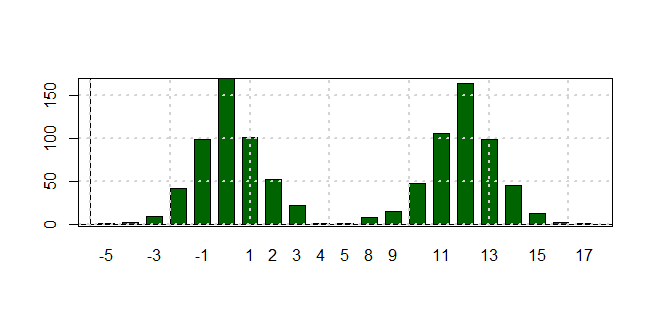}
        \label{fig:delta_bimodal}
    \end{subfigure}
    }
    \caption{ \sf \small Each row represents a   model from Table \ref{tab:Num_examples}. Left column: Scatter plot of $T_1$ ($x-$axis) and $T_2$ ($y-$axis). Right column: Bar plot of $\Delta=T_1-T_2.$ }
    \label{fig:T_and_Delta}
\end{figure}

Each model offers a distinct perspective on the relationship between the dependence structure of $\boldsymbol{T}$ and the behavior of the resulting multivariate distribution.
More importantly, each choice of $\boldsymbol{T}$ induces a different distribution for $\Delta$, as illustrated in Figure~\ref{fig:T_and_Delta}. Interestingly, even when using a simple common distribution (e.g., a Poisson distribution with parameter 1) for the components of $\boldsymbol{T}$, varying their dependence produces different shapes for the distribution of $\Delta$ and consequently different behaviors for the MDGPD.

\begin{figure}[H]
    \centering
    \begin{subfigure}{0.9\linewidth}
        \centering
        \includegraphics[width=\linewidth]{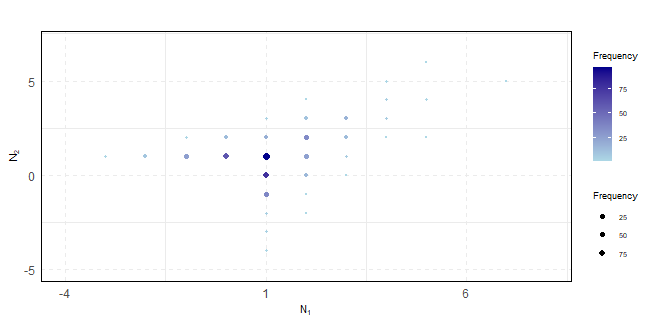}
        \caption{\sf $\boldsymbol{T}$ consists of two independent unit Poisson variables.}
        \label{fig:independent}
    \end{subfigure}\\
    \begin{subfigure}{0.9\linewidth}
        \centering
        \includegraphics[width=\linewidth]{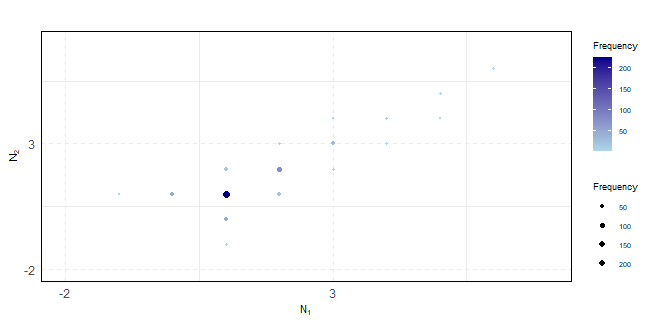}
        \caption{\sf $\boldsymbol{T}$ consists of two dependent unit Poisson variables}
        \label{fig:dependent}
    \end{subfigure}
    \\
    \begin{subfigure}{0.9\linewidth}
        \centering
        \includegraphics[width=\linewidth]{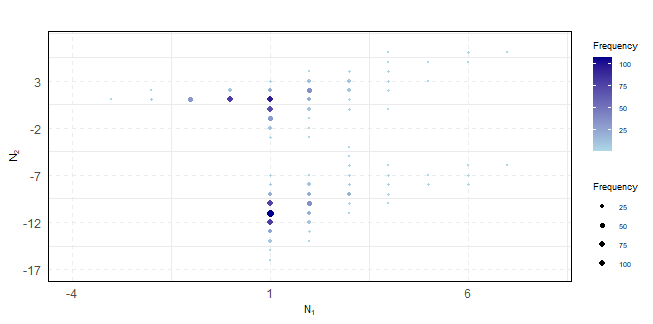}
        \caption{\sf $\boldsymbol{T}$ consists of two independent Poisson variables with   shifted by $\pm6$.}
        \label{fig:bimodal}
    \end{subfigure}
    \caption{\sf \small Bivariate discrete GPD samples obtained from Algorithm~\ref{algo:known_increments}  applied to the inputs from  Figure \ref{fig:T_and_Delta}; see also Table \ref{tab:Num_examples}.}
    \label{fig:example_intro}
\end{figure}

The goal of the bootstrap algorithm is, thus, to effectively capture the distribution of $\Delta$ in order to generate realistic simulations that reflect the dependence structure in $\boldsymbol{T}$.
\begin{table}[H]
    \centering
    \footnotesize 
    \begin{tabular}{|l|l|}
        \hline
        \textbf{Name} & \textbf{Distribution of $\boldsymbol{T}=(T_1,T_2)$} \\
        \hline
         Type (a) &  $T_1$ and $T_2$ independent with the same  unit rate  Poisson marginal\\
        \hline
        Type (b)   & Same as type (a) but with  a correlation equal to $.99$\\
        \hline
        Type (c)   & Same as type (a) but $T_1$ becomes $T_1+6$ and $T_2$ becomes $T_2-6$ \\
        \hline
    \end{tabular}
    \caption{\sf \small Three examples of  bivariate discrete generators $\boldsymbol{T}$ defined in \eqref{eq:T-max(T)}.}
    \label{tab:Num_examples}
\end{table}
In Figure~\ref{fig:T_and_Delta}, the left column of each row shows a scatter plot of random draws from $n=1000$  $\boldsymbol{T}=(T_1,T_2)$ for each model type.  
The histograms (left column) indicate how the key variable $\Delta=T_1-T_2$ changes according to the model type. Comparing the results between type (a) and type (c) highlights that the distribution of $\Delta$ can drastically vary even when the dependence remains unchanged. In these particular examples, the marginal behaviors of $T_1$ and $T_2$ are the drivers. 
In Figure~\ref{fig:example_intro}, the outputs $\boldsymbol{N}=(N_1,N_2)$ of Algorithm~\ref{algo:known_increments} are plotted from the inputs $\boldsymbol{T}=(T_1,T_2)$ derived from Figure~\ref{fig:T_and_Delta}.  
 
For some applications like the one about dry spells introduced in Figure~\ref{fig:DS_close_intro}, the practitioner may not wish to impose a parametric form on 
$\boldsymbol{T}$. To fulfill this requirement, a resampling algorithm can be proposed following the same approach as in the continuous case by \cite{legrand23}. It is described in Algorithm~\ref{algo:unknown_increments}. 
\begin{algorithm}
\caption{Bootstrap $MDGPD$ simulation}
\label{algo:unknown_increments}
\begin{algorithmic}[1]
\Require A sample of $(N_{1,i},N_{2,i})_{1\leq i\leq n}\sim MDGPD$ as in Eq. \eqref{eq:T-max(T)}
\Ensure A discrete simulated sample $(N^*_{1,k},N^*_{2,k})_{1\leq k\leq m}$ to choose 

\State Define $\Delta_i=T_{1,i}-T_{2,i}, 1\leq i\leq n,$
\State Generate $m$ generalizations $\max(\boldsymbol{N})_k\sim Geom(1-e^{-1}), 1\leq k\leq m,$ independently from $\Delta_i,$
\State Bootstrap $m$ realization $\Delta^*_k$ from $(\Delta_1,...,\Delta_n)$
\State \textbf{Return} $N_{1,k}:=\max(\boldsymbol{N})_k+\Delta^*_k\mathds{1}_{\left(\Delta^*_k<0\right)}$ and $N_{2,k}:=\max(\boldsymbol{N})_k-\Delta^*_k\mathds{1}_{\left(\Delta^*_k\geq 0\right)},$ for $1\leq k \leq m.$
\end{algorithmic}
\end{algorithm} 

\begin{figure}[H]
    \centering
    \begin{subfigure}[b]{0.5\textwidth}
        \centering
        \includegraphics[width=\textwidth]{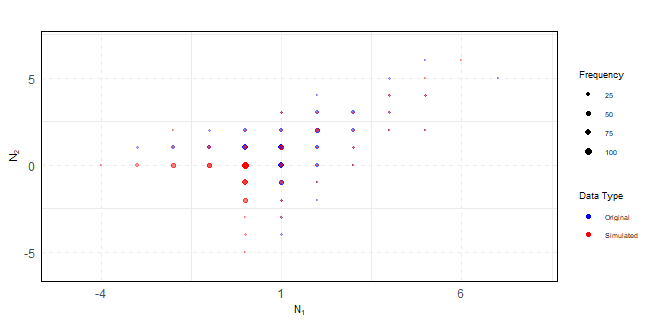}
        \label{fig:model1_image1}
    \end{subfigure}
    \hfill
    \begin{subfigure}[b]{0.49\textwidth}
        \centering
        \includegraphics[width=\textwidth]{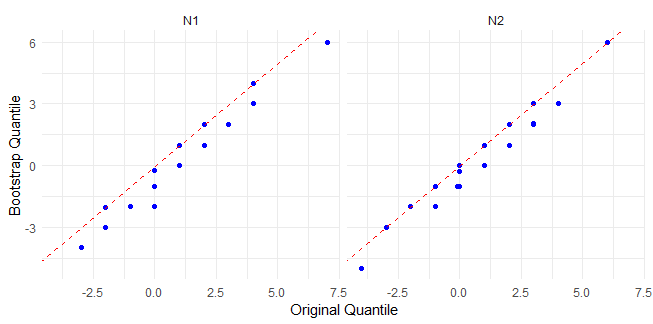}
        \label{fig:model1_image2}
    \end{subfigure}
    
    \vspace{0.2cm} 
    
    \begin{subfigure}[b]{0.5\textwidth}
        \centering
        \includegraphics[width=\textwidth]{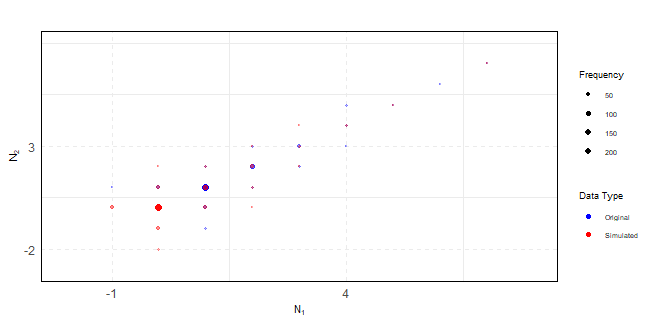}
        \label{fig:model2_image1}
    \end{subfigure}
    \hfill
    \begin{subfigure}[b]{0.49\textwidth}
        \centering
        \includegraphics[width=\textwidth]{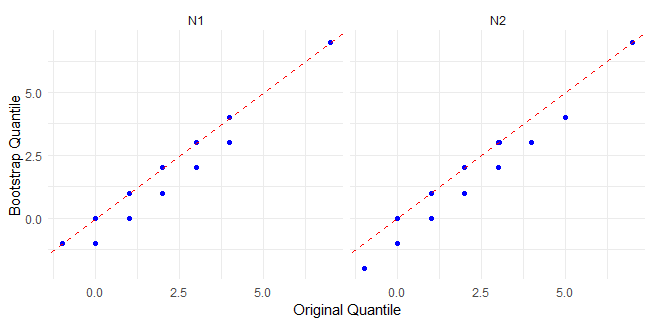}
        \label{fig:model2_image2}
    \end{subfigure}
    
    \vspace{0.2cm} 
    
    \begin{subfigure}[b]{0.5\textwidth}
        \centering
        \includegraphics[width=\textwidth]{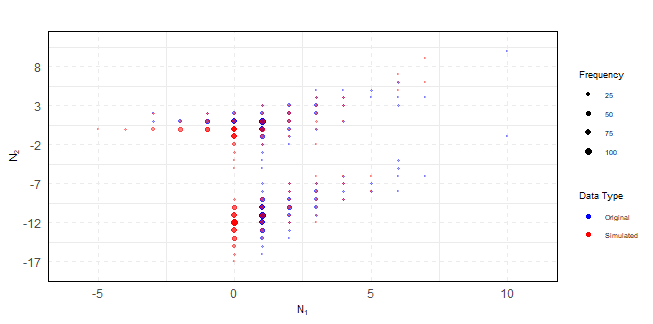}
        \label{fig:model3_image1}
    \end{subfigure}
    \hfill
    \begin{subfigure}[b]{0.49\textwidth}
        \centering
        \includegraphics[width=\textwidth]{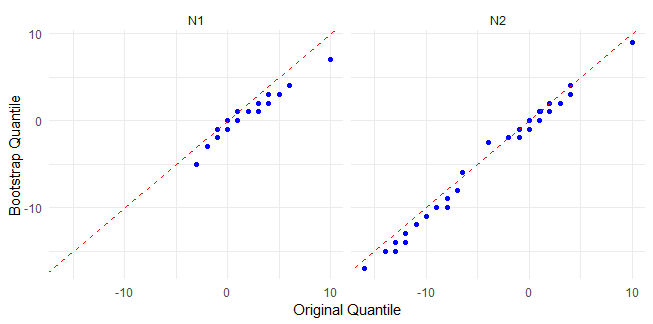}
        \label{fig:model3_image2}
    \end{subfigure}
    \caption{ \sf \small Each line corresponds from top to bottom to each of the parametric models given in Table \ref{tab:Num_examples}, in order (a) to (c). From left to right : (1) Scatter plot of simulated data with the parametric model of sample size $n=500$ (blue dots) and sampled data from one simulation using Algorithm \ref{algo:unknown_increments} with sample size $m = 500$ (red dots). The darker shades appear when multiple points overlap since there are not many possible values due to the discrete nature of the distribution. (2) Quantile-Quantile plots for $N_1$ and $N_2.$ }
    \label{fig:all_models_results}
\end{figure}

In Figure~\ref{fig:all_models_results}, we observe that the points in the Q-Q plots for all three models lie closely along the bisectrix, indicating a good fit between the distributions of the original and resampled data. This suggests that the bootstrap procedure performs well across the different parametric settings.
\\[1ex]
In summary, this section shows how to efficiently simulate a discrete bivariate Generalized Pareto Distribution (DGPD) by using the relationship $\boldsymbol{T} - \max(\boldsymbol{T}) $ and adding a geometric random variable $G $. By focusing on the bivariate case, we simplify the notation and illustrate the core algorithm for generating standard DGPD samples. The choice of distribution for the difference $\Delta = T_1 - T_2 $ plays a crucial role in shaping the resulting distribution. Three examples of $\boldsymbol{T} $ have been explored in Table~\ref{tab:Num_examples}, showing how the structure of the generator influences the behavior of $\Delta $ and subsequently the MDGPD. Figures and results highlight the impact of different generator models on the distribution of $\Delta $ and emphasize the versatility of the simulation approach. Furthermore, an extension of this method to non-parametric generators via bootstrapping has also been presented, offering flexibility in applications where the parametric form of $\boldsymbol{T} $ is unknown. The analysis suggests that the simulation technique can effectively capture the distribution of MDGPDs, both with and without dependence, allowing for broad applicability in various fields.  
\\[1ex]
The next section introduces an approach for performing inference on the bivariate DGPD. The main challenge in calibrating an MDGPD lies in its definition, where the generator can be any discrete distribution.
To address this, we adopt a simulation-based inference method that bypasses the explicit likelihood function. While the generator $\boldsymbol{T}$ still needs to be specified or estimated, this approach facilitates parameter estimation through the simulation of numerous outcomes, a task made feasible by the relative ease of simulating MDGPDs.

\section{Neural Bayes inference scheme}
\label{sec:inference}

Classical methods, such as likelihood-based estimations, being difficult to apply in our context, a promising alternative is to use machine learning methods for parameter estimation. In particular, likelihood-free parameter estimation with neural networks can offer a relevant solution. Several recent contributions use simulation-based training to estimate parameters. For instance, \cite{Lenzi2023} introduces a novel framework that leverages deep neural networks to estimate parameters in intractable models, such as max-stable processes, by training on simulations and bypassing explicit likelihood computation. Their approach demonstrates improved accuracy and efficiency compared to traditional pairwise likelihood methods, especially in high-dimensional settings. Similarly, \cite{Pacchiardi2021} proposes training generative neural networks for probabilistic forecasting by minimizing predictive-sequential scoring rules, thereby avoiding the instability of adversarial frameworks. \cite{SainsburyDale2024} presents Neural Bayes Estimators (NBEs), a class of likelihood-free methods that approximate Bayes estimators using neural networks. NBEs minimize an empirical version of the Bayes risk via Monte Carlo simulations, making them well-suited for complex models where likelihoods are intractable. Their scalability, flexibility, and compatibility with permutation-invariant architectures make them particularly attractive for replicated or high-dimensional data. Moreover, inference can be tailored by adapting the loss function used during training, offering additional flexibility. 

These likelihood-free neural estimation methods have already shown promising results on real-world environmental data, such as reanalysis data of temperature fields in the U.S. Midwest (\cite{Lenzi2023}) and sea-surface temperature extremes in the Red Sea (\cite{SainsburyDale2024}).  Motivated by these results, we adopt the methodology introduced by \cite{SainsburyDale2024}, using the publicly available implementation in Julia and R \cite{SainsburyDale2024_package}, which we adapt to our model of interest, MDGPD. Under this framework, we train a neural network on simulated data generated from the MDGPD model. This simulation-based approach allows the network to learn the dependence between parameters and observations.

As a case study, we focus on the modeling of dry spells in Switzerland. When the occurrence of dry spells at different stations exhibits spatial dependence, they form a particularly relevant application for the MDGPD framework. We simulate data that replicate the dependence patterns of such dry spells across meteorological stations, and use them to train and validate the neural estimator.

We begin by presenting the NBE methodology, followed by a description of the simulation-based training procedure and its evaluation on synthetic data. Then, we apply the methodology to Swiss dry spell data.

\subsection{Description of the NBE methodology}
\label{subsec:description_NBE}

We begin by briefly recalling the Bayesian method that is used here. Let $\theta \in \Theta \subseteq \mathbb{R}^p$ be the parameter vector of interest and $N \in \mathcal{S}$ be the observed data. The goal is to find a point estimator
\[
\hat{\theta} : \mathcal{S} \to \Theta
\]
that minimizes a loss function $L(\theta, \hat{\theta}(Z))$.

The \emph{risk function} of the estimator is defined as:
\begin{equation*}
R(\theta, \hat{\theta}) = \int_{\mathcal{S}} L(\theta, \hat{\theta}(z)) f(z|\theta)\, dz,
\end{equation*}
where $f(z|\theta)$ is the likelihood function.

The \emph{Bayes risk} is then the expectation of the risk with respect to the prior $\Pi$ on $\Theta$:
\begin{equation*}
r_\Pi(\hat{\theta}) = \int_\Theta R(\theta, \hat{\theta})\, d\Pi(\theta),
\end{equation*}
and the estimator that minimizes $r_\Pi(\hat{\theta})$ is named \emph{Bayes estimator}. 

To estimate $\theta$ on the data $N$, we consider a neural network $\hat{\theta}(N; \gamma)$ with parameter $\gamma$, the Bayes estimator of $\gamma$ being the solution of 
\[
\arg\min_\gamma r_\Pi(\hat{\theta}(N; \gamma)).
\]

Since $r_\Pi$ is typically not computable in closed form, it is approximated by Monte Carlo sampling:
\begin{equation*}
r_\Pi(\hat{\theta}(N; \gamma)) \approx \frac{1}{K} \sum_{\theta \in \vartheta} \frac{1}{J} \sum_{N \in \mathcal{N}_\theta} L(\theta, \hat{\theta}(N; \gamma)),
\end{equation*}
where $\vartheta = \{\theta^{(1)}, \ldots, \theta^{(K)}\}$ are i.i.d. samples from the prior $\Pi$, and for each $\theta^{(k)}$, $\mathcal{N}_{\theta^{(k)}} = \{N^{(1)}, \ldots, N^{(J)}\}$ are simulated datasets (with $K>1$, $J> 1$).
As in \cite{SainsburyDale2024}, we adopt the squared-error loss $L(\theta, \hat{\theta}) = \|\theta - \hat{\theta}\|^2$ in the training procedure. This choice leads the neural network to approximate the Bayes estimator under squared-error loss, which corresponds to the posterior mean. The minimization of the approximate Bayes risk is then carried out via stochastic gradient descent over simulated pairs $(\theta, N)$, without requiring explicit access to the likelihood.

When $m$ independent replicates $N_1, \dots, N_m$ of the data $N$ are available, the estimator should be invariant to permutations of the replicates. The DeepSets architecture is then used for this purpose:
\begin{align*}
\hat{\theta}(N^{(m)}; \gamma) &= \phi\left( T(N^{(m)}; \gamma_\psi); \gamma_\phi \right), \\
T(N^{(m)}; \gamma_\psi) &= a\left( \{ \psi(N_i; \gamma_\psi) : i=1,\dots,m \} \right),
\end{align*}
where 
$\psi: \mathcal{S} \to \mathbb{R}^q$ is a neural network encoding each replicate, 
$a(.)$ is a permutation-invariant aggregation function (e.g., mean), 
and $\phi: \mathbb{R}^q \to \Theta$ is a neural network decoder mapping aggregated features to parameter estimates.
In our study, the parameter of interest is $\boldsymbol{\theta} = (\boldsymbol{\sigma}, \boldsymbol{\xi},\rho,\boldsymbol{\phi})$, where 
$\boldsymbol{\sigma} \in \mathbb{R}^d$ is the scale parameter, 
$\boldsymbol{\xi} \in \mathbb{R}^d$ is the shape parameter, 
$-1<\rho<1$ the dependence parameter,
and $\boldsymbol{\phi}\in \mathbb{R}^d$ the parameters of the generator.
\\[.5ex]
The training phase of the estimator, although computationally intensive, is performed only once using simulated data. Once trained, the estimator can be rapidly applied to new datasets sharing the same structure, such as our dry spell lengths. This property makes the approach both flexible and computationally efficient in practice. Such an estimator is referred to as amortized, since the cost of training is offset by the negligible cost of inference on new data.

To illustrate how the inference method works, we first apply it to simulated data. This allows us to assess the accuracy and robustness of the approach in a controlled setting, where the true parameter values are known. Once validated on these synthetic examples, we then turn to the real-world case study: the analysis of dry spells. In this application, the observed data $N$ correspond to dry spells at two stations presented in Section~\ref{sec:intro}, which we model using the MDGPD, as introduced in Section~\ref{sec:MGDPD}.

\subsection{Simulations}
\label{ss:simulation}

Building upon the three types of generators $\boldsymbol{T}$ presented in Table~\ref{tab:Num_examples}, we now conduct a simulation study to illustrate the behavior of discrete multivariate generalized Pareto distributions (MDGPDs) under various settings.

We consider two distinct approaches scenarios for generating discrete extreme value data: 
\begin{enumerate}[label=\roman*)]
    \item[(i)] Simulation of a continuous multivariate generalized Pareto distribution (MGPD), followed by its discretization by taking the integer part;
    \item[(ii)] Direct simulation of a discrete MDGPD using the bootstrap-based Algorithm~\ref{algo:unknown_increments}.
\end{enumerate}
In both cases, we simulate 1,000 samples from each bivariate continuous and discrete GPD for a fixed value of $\lambda$, the parameter of the Poisson generator. This process is repeated for various values of $\lambda$.
\\[1ex]
{\it Scenario~(i)}. 
We first simulate samples from a continuous MGPD, as defined in Definition~\ref{def: st MGPD}, denoted $MGPD(\boldsymbol{\sigma}, \boldsymbol{\xi})$. 
In this setting, the underlying generator $\boldsymbol{T}$ is considered under all three types listed in Table~\ref{tab:Num_examples}, namely: (a) independent Poisson, (b) dependent Poisson, and (c) bimodal Poisson. 
These generator types are used consistently across both \textit{Scenario~(i)} and \textit{Scenario~(ii)} to ensure a fair comparison between the MGPD and MDGPD approaches under different dependence structures.

\begin{figure}[H]
    \centering
    \includegraphics[width=0.8\textwidth]{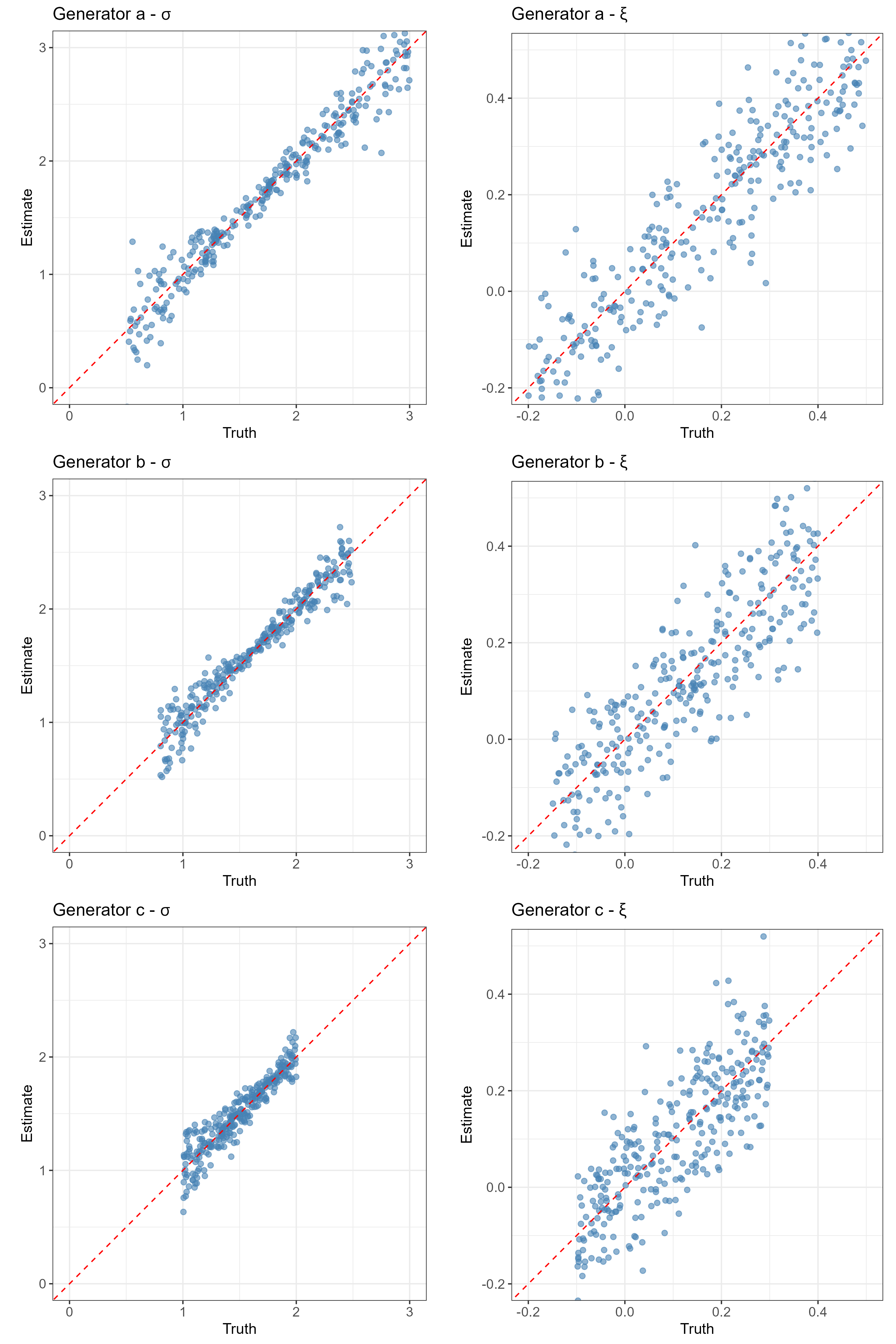}  
    \caption{\sf \small Truth versus estimate plots obtained with the Neural Bayes Estimator (NBE) for the three generator types. Each panel corresponds to a different generator and displays estimated versus true parameter values for $\boldsymbol{\sigma}$ (left column) and $\boldsymbol{\xi}$ (right column).}
    \label{fig:NBE_on_sim}
\end{figure}

Following the standardization step (where the maximum component is exponentially distributed with rate $1$), a transformation is applied to introduce non-standard margins, by specifying scale parameters $\boldsymbol{\sigma}$ and shape parameters $\boldsymbol{\xi}$.
The continuous samples are then discretized by taking the integer part of each component, mimicking practical way of doing when observed data are discrete (e.g., counts, rounded measurements).
\\[1ex]
{\it Scenario~(ii)}. 
We simulate samples directly from a discrete MDGPD, using Algorithm~\ref{algo:known_increments} with the same generator type (a) for consistency. 
This approach captures the inherent discreteness of the model without requiring any post-processing.
\\[1ex]
{\it Comparison between the two scenarios}.
The estimation performance and goodness-of-fit are compared between the two data generation approaches, considering different configurations of the true scale ($\boldsymbol{\sigma}$) and shape ($\boldsymbol{\xi}$) parameters. 
For each setting, Table~\ref{tab:results_sim} reports the true parameter values and their corresponding estimates obtained from fitting either the discretized MGPD or the MDGPD. Here, considering the results across all generator types, we observe that:
\begin{itemize}
    \item The MDGPD approach yields slightly more accurate parameter estimates, particularly for the scale parameter $\sigma$;
    \item The log-likelihood values are systematically higher (i.e., less negative) for the MDGPD, indicating a better model fit;
    \item The Kolmogorov–Smirnov (KS) distances are lower for the MDGPD, suggesting closer adherence to the empirical distributions.
\end{itemize}
These findings confirm that modeling the data through a discrete MDGPD, rather than relying on discretized continuous models, provides significant advantages in terms of both parameter recovery and fit quality.
\\[1ex]
The simulation results also emphasize the critical role of the generator $\boldsymbol{T}$ in shaping the dependence structure of the MDGPD, which offers flexibility in terms of models, even if a challenge on real data.
\\[1ex]
Overall, the simulation study highlights the relevance of discrete modeling approaches for extreme value analysis when dealing with inherently discrete data. 
The MDGPD framework improves performance compared to discretized MGPDs, particularly in capturing tail behaviors and dependence structures.
\\[1ex]
Figure~\ref{fig:NBE_on_sim} displays the truth versus estimate plots for the scale parameter $\boldsymbol{\sigma}$ and the shape parameter $\boldsymbol{\xi}$ across the three types of generators introduced earlier. 
The estimation of $\boldsymbol{\sigma}$ appears satisfactory, with points closely aligned along the identity line, whereas the estimation of $\boldsymbol{\xi}$ exhibits greater variability, particularly for extreme values.

This phenomenon is consistent with the well-known challenges encountered in inference for the GPD. 
The estimation of the shape parameter $\xi$ is notably more difficult than that of the scale parameter $\sigma$ for several reasons: (i) $\xi$ governs the tail heaviness of the distribution, making small estimation errors particularly impactful for extrapolations to rare events (see e.g. \cite{Coles2001}); (ii) the Fisher information associated with $\xi$ can degenerate near critical values, leading to increased estimator variance (\cite{deHaanFerreira2006}); and (iii) estimators of $\xi$ generally exhibit slower asymptotic convergence rates compared to those for $\sigma$ (\cite{Pickands1975}).

Thus, the increased dispersion observed in the estimates of $\xi$ is not an artifact of the model, but rather an intrinsic difficulty inherent to tail modeling.

\begin{table}[H]
\centering
\caption{\sf \small Comparison of MDGPD and Discretized-MGPD Parameter Estimates across Generator Types with Log-likelihood and KS statistic (in \textbf{bold} for the MDGPD)}
\label{tab:results_sim}
\small
\resizebox{\textwidth}{!}{%
\begin{tabular}{llcccccc}
\toprule
\textbf{Generator} & \textbf{Parameter} & \textbf{True} & \textbf{Discretized} & \textbf{MDGPD} & \textbf{Rel. Error (\%)} & \textbf{Log} & \textbf{KS} \\
\textbf{Type}& &  &- \textbf{MGPD}&&&\textbf{-Likelihood}& \textbf{Statistic}
\\
\midrule
\multirow{7}{*}{(a) Poisson Indep.} 
 & $\sigma_1$   & 1.5 & 1.45 & 1.50 & 3.33  & \multirow{7}{*}{-235.6 / \textbf{-225.4}} & \multirow{7}{*}{0.14 / \textbf{0.07}} \\
 & $\sigma_2$   & 1.5 & 1.40 & 1.52 & 6.67  & & \\
 & $\xi_1$      & 0.2 & 0.17 & 0.19 & 15.00 & & \\
 & $\xi_2$      & 0.2 & 0.15 & 0.18 & 25.00 & & \\
 & $\rho$       & 0.0 & —    & 0.02 & —     & & \\
 & $\lambda_1$  & 1.0 & —    & 1.05 & —     & & \\
 & $\lambda_2$  & 1.0 & —    & 0.98 & —     & & \\
\midrule
\multirow{7}{*}{(b) Poisson Dep.} 
 & $\sigma_1$   & 2.0 & 1.92 & 2.05 & 4.00  & \multirow{7}{*}{-180.3 / \textbf{-175.1}} & \multirow{7}{*}{0.21 / \textbf{0.13}} \\
 & $\sigma_2$   & 2.0 & 1.95 & 2.02 & 2.50  & & \\
 & $\xi_1$      & -0.1& -0.15& -0.08& 50.00 & & \\
 & $\xi_2$      & -0.1& -0.13& -0.09& 30.00 & & \\
 & $\rho$       & 0.9 & —    & 0.88 & —     & & \\
 & $\lambda_1$  & 1.0 & —    & 1.02 & —     & & \\
 & $\lambda_2$  & 1.0 & —    & 1.01 & —     & & \\
\midrule
\multirow{7}{*}{(c) Poisson Bimodal} 
 & $\sigma_1$   & 2.5 & 2.35 & 2.55 & 6.00  & \multirow{7}{*}{-198.9 / \textbf{-192.4}} & \multirow{7}{*}{0.19 / \textbf{0.12}} \\
 & $\sigma_2$   & 2.5 & 2.30 & 2.48 & 8.00  & & \\
 & $\xi_1$      & 0.1 & 0.05 & 0.09 & 50.00 & & \\
 & $\xi_2$      & 0.1 & 0.04 & 0.08 & 60.00 & & \\
 & $\rho$       & 0.0 & —    & 0.01 & —     & & \\
 & $\lambda_1$  & 1.0 & —    & 1.10 & —     & & \\
 & $\lambda_2$  & 1.0 & —    & 0.90 & —     & & \\
\bottomrule
\end{tabular}
}
\end{table}

\subsection{Dry spells}

In the literature, many studies have explored extreme dry spell durations, primarily employing traditional continuous BM-GEV and POT-GP methods (\cite{Lana2006, Serra2013,Cindri2018}). Now that we used the inference scheme on simulations to find the parameters on different types of MDGPDs, we turn to its application for calibrating MDGPDs on dry spells. \\ A reproducible plan for analyzing dry spells is given below (which code R can be provided upon request).
\begin{enumerate}
    \item \textbf{Data preparation and thresholding.} Select meteorological stations of interest (e.g., Interlaken and Lauterbrunnen in this example). From daily precipitation records, compute dry spell lengths defined as the number of consecutive days with rainfall below a fixed threshold (e.g., 1 mm). Focus on extreme events by selecting dry spells exceeding the 99\% quantile station-wise.

    \item \textbf{Empirical analysis.} Compute the empirical differences $\Delta = N_1 - N_2$ between station-wise dry spell lengths and assess their distribution.

    \item \textbf{Selection of generator.} Based on the structure of $\Delta$, select an appropriate generator $\boldsymbol{T}$. 
    \item \textbf{Neural inference.} Apply the trained estimator to the dry spell observations. Parameter uncertainty is quantified via parametric bootstrap, yielding confidence intervals for $(\boldsymbol{\hat{\sigma}}, \boldsymbol{\hat{\xi}}, \hat{\rho},\boldsymbol{\hat{\phi}})$.

    \item \textbf{Model validation.} Assess the fit of the MDGPD model using:
    \begin{itemize}
        \item scatter plots comparing observed and simulated dry spell pairs,
        \item marginal Q-Q plots against bootstrap samples,
        \item Root Mean Squared Errors (RMSE) metrics.
    \end{itemize}
\end{enumerate}

\begin{figure}[H]
    \centering
    \includegraphics[width=0.8\linewidth]{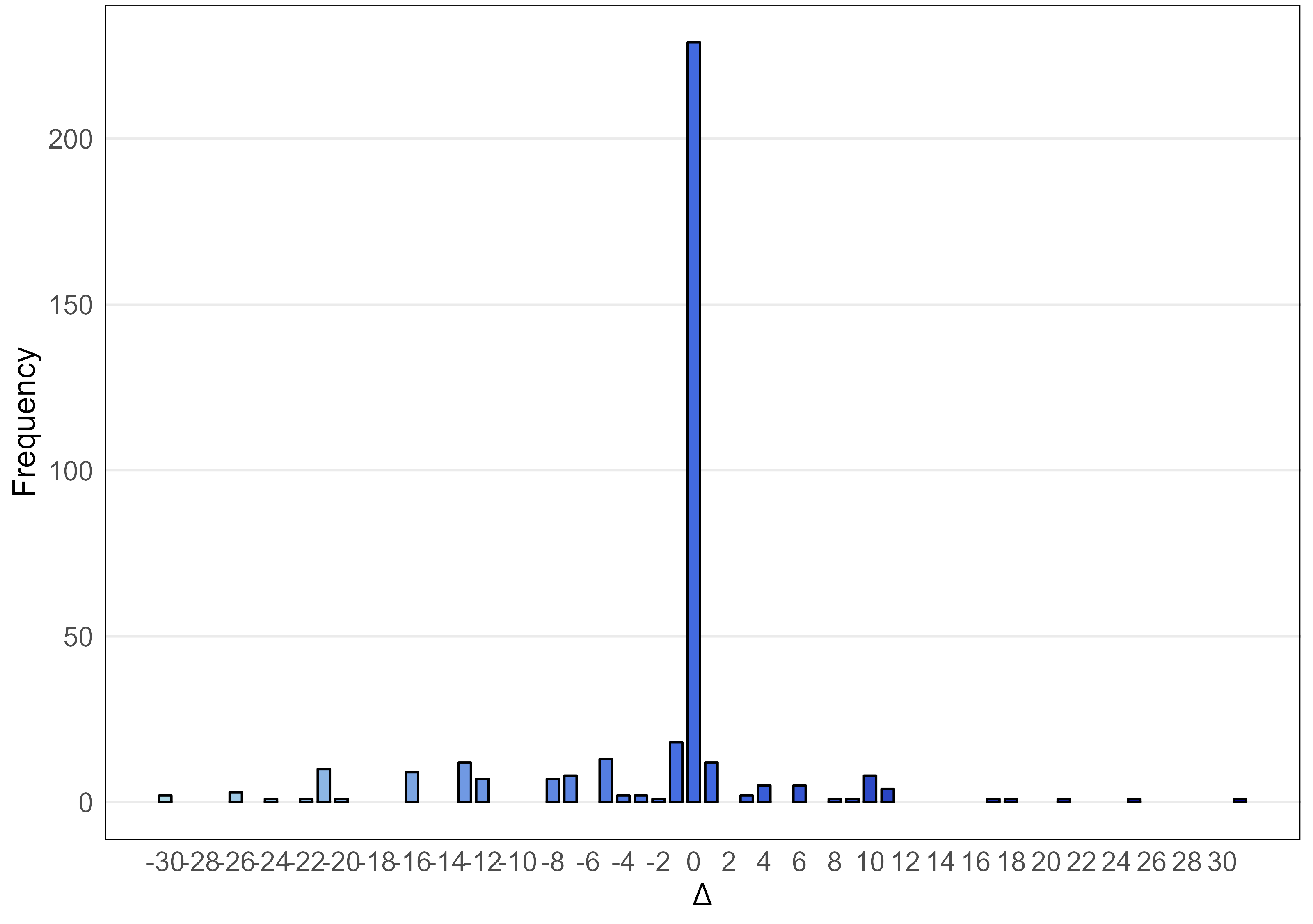}
    \caption{\sf \small Observation of $\Delta$ for the two stations Lauterbrunnen and Interlaken}
    \label{fig:Delta_close}
\end{figure}

We focus on two stations from a precipitation dataset covering 105 locations in Switzerland, introduced by \cite{Pasche_2022}, which contains over 31,000 daily observations. Rather than analyzing precipitation amounts, we use the dataset to study dry spells. Specifically, we compute the lengths of dry spells for the two selected stations, following the procedure described above. Then, we compute the empirical differences $\Delta = N_1 - N_2$, as illustrated in Figure~\ref{fig:Delta_close}.

From Figure~\ref{fig:Delta_close}, we observe that $\Delta$ is centered around $0$, with a clear mode at $0$ and very few values elsewhere on either side. Among the three types of generators studied in Table~\ref{tab:Num_examples}, this behavior is most similar to that of type~(b). Based on this observation, we apply the NBE method, previously trained, to the dry spell data.

For this application, the fitted MDGPD model yields estimates of the marginal parameters $\hat{\sigma}_1$, $\hat{\sigma}_2$, $\hat{\xi}_1$, and $\hat{\xi}_2$, as well as the dependence parameter $\hat{\rho}$ and generator parameters $\hat{\lambda}_1$, $\hat{\lambda}_2$, as summarized in Table~\ref{tab:metric_dist}. Confidence intervals at the $95\%$ level are computed via parametric bootstrap, and estimation uncertainty is quantified through the RMSE of the bootstrap replicates.

\begin{figure}[ht]
    \centering
    \begin{subfigure}[b]{0.49\textwidth}
        \centering
        \includegraphics[width=\textwidth]{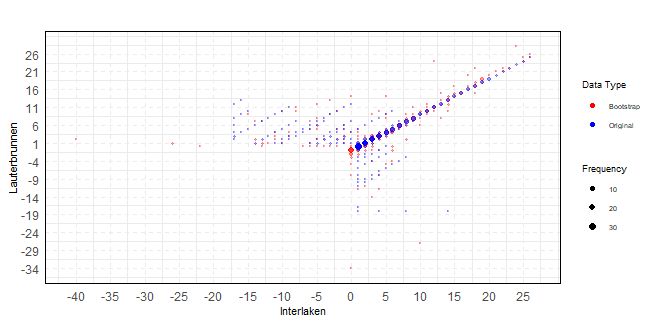}
        \label{fig:scatter_close}
    \end{subfigure}
    \hfill
    \begin{subfigure}[b]{0.49\textwidth}
        \centering
        \includegraphics[width=\textwidth]{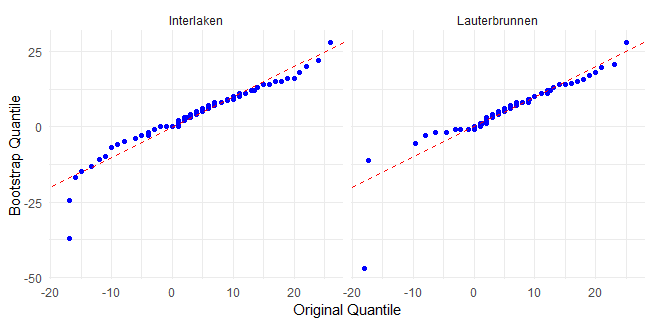}
        \label{fig:qq_close}
    \end{subfigure}
    \caption{\sf \small Left column: Scatter plot of dry spells at two close stations and bootstrap samples from the fitted model (blue: observed, red: bootstrap). Right column: Quantile-Quantile plot of observed dry spells and bootstrap samples.  The fit captures the overall pattern well, particularly for the right tail, our interest for this analysis.}
    \label{fig:close_fit}
\end{figure}

\begin{table}[H]
\centering
\caption{\sf \small Analysis of dry spells from Figure~\ref{fig:DS_close_intro} with fitting metrics RMSE}
\label{tab:metric_dist}
\begin{tabular}{l|ccc}
\toprule
\textbf{Parameter} & \textbf{Estimate} & \textbf{CI ($95\%$)} & \textbf{RMSE} \\
\hline
$\sigma_1$ & 7.85 & $[7.10,\;8.60]$ & 0.62 \\
$\sigma_2$ & 9.12 & $[8.30,\;10.05]$ & 0.74 \\
$\xi_1$    & -0.18 & $[-0.26,\;-0.10]$ & 0.09 \\
$\xi_2$    & -0.25 & $[-0.33,\;-0.17]$ & 0.12 \\
$\rho$     & 0.82  & $[0.74,\;0.89]$ & 0.04 \\
$\lambda_1$ & 1.05 & $[0.92,\;1.18]$ & 0.08 \\
$\lambda_2$ & 1.10 & $[0.97,\;1.24]$ & 0.09 \\
\bottomrule
\end{tabular}
\end{table}
The RMSE values in Table~\ref{tab:metric_dist} show that the parameter estimates are reasonably accurate. The error remains small for both $\sigma$ and $\xi$ (comparable with the RMSE obtained in the simulation study; see Table~\ref{tab:results_sim}), which supports the good visual fit observed in the Q-Q plot (Figure~\ref{fig:close_fit}). This confirms that the model captures well the distribution of extreme dry spell lengths.

This approximation yields satisfactory results. However, it is important to note that the present application serves primarily as an illustration. A comprehensive evaluation would require testing a broader range of model families for the generator $\boldsymbol{T}$ (as done, for instance, in the continuous setting by \cite{Kiriliouk2019})), or developing alternative approaches. We have opted for the latter, which we are currently investigating and plan to address more thoroughly in future work.

\section{Conclusion}

In this work, we have proposed a new discrete multivariate framework for modeling extreme events, based on the construction of a MDGPD. This approach allows for a more accurate representation of discrete extreme data, such as dry spells, and offers a flexible simulation scheme that captures complex dependence structures.
To perform inference on the MDGPD parameters, we have adapted a NBE method, relying on likelihood-free inference through simulations. The simulation-based training of the neural network enables the bypassing of explicit likelihood evaluation, offering an efficient and scalable solution.
Alternative inference strategies could also be explored. In particular, 
a promising direction for future work, which is the focus of our ongoing research, involves leveraging optimal transport methods (see, e.g., \cite{flamary2021pot, singha2024comparingmultivariatedistributionsnovel}) to perform inference by learning mappings between discrete distributions.
Optimal transport offers a principled way to measure distances between probability distributions and could facilitate efficient matching between observed data and simulated models without relying on likelihood evaluation. By transferring mass from one discrete distribution to another in a cost-optimal way, this strategy may overcome some challenges inherent in discrete modeling and accelerate the estimation process.

\subsection*{Acknowledgments}
The work of S. Aka was supported by the Agence Nationale Recherche Technologie (ANRT) with the CIFRE PhD funding.
Part of Naveau’s research work was supported by European H2020 XAIDA (Grant agreement ID: 101003469) and the  French national programs:  the Agence Nationale de la Recherche (ANR) EXSTA, the PEPR TRACCS programme under grant number  EXTENDING (ANR-22-EXTR-0005),  
the PEPR   IRIMONT, (France 2030 ANR-22-EXIR-0003), and the SHARE PEPR Maths-Vives.

\newpage
\appendix
\section{Appendix}
\label{app:proofs}

\subsection{Proof of Proposition~\ref{prop:Z_over_Gamma}}\label{App:prop1}


Let $\Lambda$ follow a $Gamma(\alpha, \beta)$ distribution. Then, for $Z$ following a $MGPD(\boldsymbol{1},\boldsymbol{0},\boldsymbol{S})$ and independent of $\Lambda,$ we can write 
\begin{align*}
\mathbb{P}\left(\frac{\boldsymbol{Z}}{\Lambda}\leq \boldsymbol{x}\right)&=\mathbb{P}\left(\boldsymbol{S}+E\leq \boldsymbol{x}\Lambda\right), \text{ by definition of the $MDGPD$,}\\
    &=1-\mathbb{P}\left(E> \min(\boldsymbol{x}\Lambda-\boldsymbol{S})\right), \\
    &=1-\int_0^{+\infty}\mathbb{P}\left(E> \min(\boldsymbol{x}y-\boldsymbol{S})\right)\frac{\beta^{\alpha}y^{\alpha-1}}{\Gamma(\alpha)}e^{-\beta y}dy,\\
    &=1-\int_0^{+\infty}\mathbb{E}\left(1\wedge  e^{\max(\mathbf{S}-\boldsymbol{x}y)}\right)\frac{\beta^{\alpha}y^{\alpha-1}}{\Gamma(\alpha)}e^{-\beta y}dy,\\
    &=1-\mathbb{E}\left(1\wedge \int_0^{+\infty}e^{\max(\mathbf{S}-\boldsymbol{x}y)}\frac{\beta^{\alpha}y^{\alpha-1}}{\Gamma(\alpha)}e^{-\beta y}dy\right), \text{ by Fubini theorem,}\\
    &=1-\mathbb{E}\left(1\wedge\max\left(\int_0^{+\infty}e^{\mathbf{S}-\boldsymbol{x}y}\frac{\beta^{\alpha}y^{\alpha-1}}{\Gamma(\alpha)}e^{-\beta y}dy\right)\right), \text{ by monotonicity of the exponential,}\\
    &=1-\mathbb{E}\left(1\wedge\max\left(e^{\mathbf{S}}\int_0^{+\infty}e^{-\boldsymbol{x}y}\frac{\beta^{\alpha}y^{\alpha-1}}{\Gamma(\alpha)}e^{-\beta y}dy\right)\right),\\
    &=1-\mathbb{E}\left(1\wedge e^{\max(\mathbf{S}-\alpha \log(1+\frac{\mathbf{x}}{\beta}))}\right).
\end{align*}

\noindent So $\boldsymbol{Z}/\Lambda$ follows a $MGPD(\boldsymbol{\frac{\beta}{\alpha}},\boldsymbol{\frac{1}{\alpha}},\boldsymbol{S})$. \hfill $\Box$

\subsection{Proofs for Section~2 }
\label{app:sec2}

\begin{proof}[\textbf{Proof of Equation \eqref{eq:max_Z}}]

By definition of the spectral vector, we know that $\max(\boldsymbol{Z})=0,$ which means that there exists $j\in \{1,\ldots,d\}$  such that $S_j=0$ and   $\max(\boldsymbol{Z})=\max(\boldsymbol{S}+E)=E.$ Therefore, we can write
\begin{align*} 
  \max(\lceil\boldsymbol{Z}\rceil)&=  \max(\lceil\boldsymbol{S}+E\rceil)\text{, by Definition \ref{th_Z+E_Rootzen_et_al.}},\\
  &=\underset{i=1,\ldots,d}{\max}(\lceil S_i+E\rceil)\text{, since $\boldsymbol{S} \in \mathbb{R}^d,$}\\
  &=\lceil \underset{i=1,\ldots,d}{\max}S_i+E\rceil\text{, from the properties of the integer part,}\\
  &=\lceil 0 +E\rceil\text{, since $\max(\boldsymbol{S})=0$,}\\
  &=\lceil E\rceil.
\end{align*} 
\end{proof}

\begin{proof}[\textbf{Proof of Corollary~\ref{cor:N_marg}}]~

$\bullet$ \textit{Marginals computation.} Let $ \Delta = N_1 - N_2 $ and $ G = \max(N_1, N_2) $. By assumption, $ \Delta $ and $ G $ are independent, with
\[
\mathbb{P}(\Delta = l, G = g) = \mathbb{P}(\Delta = l)\, \mathbb{P}(G = g) = \mathbb{P}(\Delta = l)\, e^{-(g - 1)}(1 - e^{-1}).
\]
We express $ N_1 $ and $ N_2 $ as functions of $ \Delta $ and $ G $:
\[
\begin{cases}
N_1 = G + \Delta - \max(0, \Delta), \\
N_2 = G - \max(0, \Delta).
\end{cases}
\]
This implies that for any $(n_1, n_2)$ with $ \max(n_1, n_2) \geq 1 $,
\[
\mathbb{P}(N_1 = n_1, N_2 = n_2) = e^{-(\max(n_1, n_2) - 1)}(1 - e^{-1}) \, \mathbb{P}(\Delta = n_1 - n_2).
\]

Let us now compute the marginal laws for any $n \in \mathbb{Z}$.

For $n<1$, we have
\begin{align*}
    \mathbb{P}(N_1=n)&=\sum_{k=1}^{\infty}\mathbb{P}(N_1=n,N_2=k)\\
    &=\sum_{k=1}^{\infty}e^{-(\max(n,k)-1)}(1-e^{-1})\mathbb{P}\left(\Delta=n-k\right)\\
    &=(1-e^{-1})e\sum_{k=1}^{\infty}e^{-k}\mathbb{P}\left(\Delta=n-k\right)=(1-e^{-1})e\sum_{l=-\infty}^{n-1}e^{-(n-l)}\mathbb{P}\left(\Delta=l\right)\\
    &=(1-e^{-1})e^{-(n-1)}\mathbb{E}\left(e^{\Delta}\mathds{1}_{(\Delta <n)}\right).
\end{align*}

For $n\geq 1$,
\begin{align*}
    \mathbb{P}(N_1=n)&=(1-e^{-1})e\sum_{k=-\infty}^{n}e^{-n}\mathbb{P}\left(\Delta=n-k\right) + (1-e^{-1})e\sum_{k=n+1}^{\infty}e^{-k}\mathbb{P}\left(\Delta=n-k\right)\\
    &=(1-e^{-1})e^{-(n-1)}\sum_{l\geq 0}\mathbb{P}\left(\Delta=l\right)+(1-e^{-1})e\sum_{l=-\infty}^{-1}e^{-(n-l)}\mathbb{P}\left(\Delta=l\right)\\
    &=(1-e^{-1})e^{-(n-1)}\left[\mathbb{P}\left(\Delta\geq 0\right)+\mathbb{E}\left(e^{\Delta}\mathds{1}_{(\Delta < 0)}\right)\right].
\end{align*}
In the same way, for $n<1,$
\begin{align*}
    \mathbb{P}(N_2=n)&=\sum_{k=1}^{\infty}\mathbb{P}(N_1=k,N_2=n)\\
    &=\sum_{k=1}^{\infty}e^{-(\max(k,n)-1)}(1-e^{-1})\mathbb{P}\left(\Delta=k-n\right)\\
    &=(1-e^{-1})e\sum_{k=1}^{\infty}e^{-k}\mathbb{P}\left(\Delta=k-n\right)=(1-e^{-1})e\sum_{l=1-n}^{+\infty}e^{-(n+l)}\mathbb{P}\left(\Delta=l\right)\\
    &=(1-e^{-1})e^{-(n-1)}\mathbb{E}\left(e^{-\Delta}\mathds{1}_{(\Delta\geq 1-n)}\right).
 \end{align*}
And, for $n\geq 1,$

\begin{align*}
    \mathbb{P}(N_2=n)&=(1-e^{-1})e\sum_{k=-\infty}^{n}e^{-n}\mathbb{P}\left(\Delta=k-n\right) + (1-e^{-1})e\sum_{k=n+1}^{\infty}e^{-k}\mathbb{P}\left(\Delta=k-n\right)\\
    &=(1-e^{-1})e^{-(n-1)}\sum_{l=-\infty}^0 \mathbb{P}\left(\Delta=l\right)+(1-e^{-1})e\sum_{l=1}^{+\infty}e^{-(n+l)}\mathbb{P}\left(\Delta=l\right)\\
    &=(1-e^{-1})e^{-(n-1)}\left[\mathbb{P}\left(\Delta\leq 0\right)+\mathbb{E}\left(e^{-\Delta}\mathds{1}_{(\Delta > 0)}\right)\right].
\end{align*}

$\bullet$ \textit{Survival functions.}  Instead of computing the survival cdf from the marginals obtained above, we choose to compute it directly using conditional expectation; this helps reduce the computation.

Since $N_1 = G + \Delta\mathds{1}_{(\Delta < 0)}$, we derive:
\begin{align*}
\mathbb{P}(N_1 > n) 
 &= \mathbb{E}\left(\mathds{1}_{\left(G>n-\Delta\mathds{1}_{\left(\Delta<0\right)}\right)}\right)\\
&=\mathbb{E}\left(\mathbb{E}\left(\mathds{1}_{\left(G>n-\Delta\mathds{1}_{\left(\Delta<0\right)}\right)}|\Delta<0\right)\right)+\mathbb{E}\left(\mathbb{E}\left(\mathds{1}_{\left(G>n-\Delta\mathds{1}_{\left(\Delta<0\right)}\right)}|\Delta\geq 0\right)\right)\\
&= \mathbb{E}\left( \mathbb{P}\left(G > n - \Delta\,\middle|\, \Delta<0 \right) \right) +\mathbb{E}\left( \mathbb{P}\left(G > n\,\middle|\, \Delta\geq0 \right) \right) \\
&= \mathbb{E}\left( 
    \begin{cases}
        1, & \text{if } n\leq \Delta< 0 \text{ and } n< 0 \\
        e^{-(n - \Delta )}, & \text{if } \Delta\leq n- 1 \text{ and } n< 1
    \end{cases}
\right)+ \mathbb{E}\left(
    \begin{cases}
        1, & \text{if } n < 1 \text{ and }\Delta\geq 0 \\
        e^{-n}, & \text{if } n\geq 1 \text{ and }\Delta\geq 0 
    \end{cases}
\right)\\
&= \mathbb{E}\left(\mathds{1}_{(n\leq\Delta<0)}\mathds{1}_{(n<0)}\right)+\mathbb{E}\left( e^{-(n - \Delta )}\mathds{1}_{(\Delta\leq n-1)}\mathds{1}_{(n<1)}\right)\\&+\mathbb{E}\left(\mathds{1}_{(n<1)}\mathds{1}_{(\Delta\geq 0)}\right)+e^{-n}\mathbb{E}\left(\mathds{1}_{(n\geq1)}\mathds{1}_{(\Delta\geq 0)}\right),
\end{align*}
hence,
\begin{equation}\label{eq:cdfN1}
   \mathbb{P}(N_1 > n) =\mathbb{P}\left(n\leq\Delta<0\right)\mathds{1}_{(n<0)}+e^{-n}\mathbb{E}\left( e^{\Delta}\mathds{1}_{(\Delta\leq n-1)}\right)\mathds{1}_{(n\leq0)}+\mathbb{P}\left(\Delta\geq 0\right)\left(\mathds{1}_{(n\leq0)}+e^{-n}\mathds{1}_{(n\geq 1)}\right).
\end{equation}
Similarly, noting that $N_2=G-\Delta\mathds{1}_{(\Delta\geq0)}=G-\Delta\mathds{1}_{(\Delta>0)},$
\begin{equation}\label{eq: P(N2>n)}
 \mathbb{P}(N_2 > n) = \mathbb{P}\left(0<\Delta\leq-n\right)\mathds{1}_{(n<0)}+e^{-n}\mathbb{E}\left( e^{-\Delta}\mathds{1}_{(\Delta\geq 1-n)}\right)\mathds{1}_{(n\leq0)}+\mathbb{P}\left(\Delta\leq 0\right)\left(\mathds{1}_{(n\leq 0)}+e^{-n}\mathds{1}_{(n\geq 1)}\right).   
\end{equation}

$\bullet$ \textit{Conditional law.}  
Conditioned on $N_1 > 0$, $N_1$ has the same law as $G$,
$$
\mathbb{P}(N_1 > n \mid N_1 > 0) 
= \mathbb{P}(G > n - \Delta \mathds{1}_{(\Delta < 0)} \mid G > -\Delta \mathds{1}_{(\Delta < 0)}) \\
= \mathbb{P}(G > n),
$$
by the memoryless property of the geometric distribution.\\[1ex]
$\bullet$ \textit{Extension to dimension $d > 2$.} 
Let $\boldsymbol{T} = (T_1, \ldots, T_d)$ be a vector of discrete random variables, and define for each $i \in \{1, \ldots, d\}$:
\[
\Delta_i := T_i - \max_{j \neq i} T_j, \quad G := \max(N_1, \ldots, N_d), \quad N_i := G + \Delta_i \mathds{1}_{(\Delta_i < 0)}.
\]
For any $d>2,$ the expressions of $\mathbb{P}\left(N_i>n\right)$ and $\mathbb{P}\left(N_i>n|N_i>0\right)$ are the same as those given for $N_1$. Therefore, for any $n\in \mathbb{Z},$
\begin{equation*}
    \mathbb{P}\left(N_i> n\right)= \mathbb{P}\left(n\leq\Delta_i<0\right)\mathds{1}_{(n<0)}+e^{-n}\mathbb{E}\left( e^{\Delta_i}\mathds{1}_{(\Delta_i\leq n-1)}\right)\mathds{1}_{(n\leq0)}+\mathbb{P}\left(\Delta_i\geq 0\right)\left(\mathds{1}_{(n\leq0)}+e^{-n}\mathds{1}_{(n\geq 1)}\right),
\end{equation*}
and 
\begin{equation*}
   \mathbb{P}(N_i > n \mid N_i > 0) =\mathbb{P}(G> n). 
\end{equation*}

These general formulas also hold for $N_2$, once having noticed that
$N_2=G-\Delta\mathds{1}_{(\Delta\geq0)}$ can be written as $N_2=G+\Delta_2\mathds{1}_{(\Delta_2<0)}$, with $\Delta_2=T_2-T_1=-\Delta$. We used initially, for $d=2$ (see Equation~\eqref{eq: P(N2>n)}), the random variable $\Delta$, common to both $N_1$ and $N_2$, to simplify the computation.
\end{proof}

\begin{proof}[\textbf{Proof of Corollary ~\ref{prop:Dist_AZ}}]\label{pr:Dist_AN}
Note that, for $\boldsymbol{k}\in \mathbb{Z}^d,$
\begin{align*}
    \boldsymbol{AN} \nleqslant \boldsymbol{m}  
    &\iff \exists i\in \{1,\dots,n\},\ \sum_{j=1}^d a_{ij} (S_j+G) > m_i \quad \text{(by definition of } \boldsymbol{N}\text{)} \\
    &\iff \exists i\in \{1,\dots,n\},\ G > \frac{m_i - \sum_{j=1}^d a_{ij} S_j}{\sum_{j=1}^d a_{ij}}.
\end{align*}
Since $G$ is independent of $\boldsymbol{S}$ and follows a geometric distribution with parameter $1-e^{-1},$ we obtain
\begin{align}
    \mathbb{P}(\boldsymbol{AN} \nleqslant \boldsymbol{m})&=\mathbb{E}\left[1\wedge \exp\left\{-\left(\left\lceil\underset{i=1,...,n}{\min}\frac{m_i-\sum_{j=1}^d a_{ij}S_j}{\sum_{j=1}^d a_{ij}}\right\rceil\right)\right\}\right], \notag \\
    &=\mathbb{E}\left[1\wedge \exp\left\{-\left(\left\lceil-\underset{i=1,...,n}{\max}\frac{\sum_{j=1}^d a_{ij}S_j-m_i}{\sum_{j=1}^d a_{ij}}\right\rceil\right)\right\}\right], \notag \\
    &=\mathbb{E}\left[1\wedge \exp\left\{\left\lceil\underset{i=1,...,n}{\max}\frac{\sum_{j=1}^d a_{ij}S_j-m_i}{\sum_{j=1}^d a_{ij}}\right\rceil\right\}\right].\label{eq:AN nleqant n}
\end{align}

So,

\begin{align} \label{eq_AN_nleqslant_0}
  \mathbb{P}(\boldsymbol{AN-m} \leq \boldsymbol{k}|\boldsymbol{AN} \nleqslant \boldsymbol{m})&= \frac{\mathbb{P}(\boldsymbol{AN-m} \leq \boldsymbol{k},\boldsymbol{AN} \nleqslant \boldsymbol{m})}{\mathbb{P}(\boldsymbol{AN} \nleqslant \boldsymbol{m})} \nonumber\\
  &=\frac{\mathbb{P}(\boldsymbol{AN-m} \leq \boldsymbol{k})-\mathbb{P}(\boldsymbol{AN-m} \leq \boldsymbol{k\wedge \boldsymbol{0}})}{\mathbb{P}(\boldsymbol{AN} \nleqslant \boldsymbol{m})}
\nonumber\\
&=\frac{\mathbb{P}(\boldsymbol{AN-m} \nleqslant \boldsymbol{k}\wedge \boldsymbol{0})-\mathbb{P}(\boldsymbol{AN-m} \nleqslant \boldsymbol{k})}{\mathbb{P}(\boldsymbol{AN} \nleqslant \boldsymbol{m})}.
\end{align}
Then, using \eqref{eq:AN nleqant n}, we obtain
\begin{align*}
   &\mathbb{P}(\boldsymbol{AN-m} \nleqslant \boldsymbol{k}\wedge \boldsymbol{0})-\mathbb{P}(\boldsymbol{AN-m} \nleqslant \boldsymbol{k})=\\
   &\mathbb{E}\left[1\wedge \exp\left\{\left\lceil\underset{i=1,...,n}{\max}\frac{\sum_{j=1}^d a_{ij}S_j-m_i}{\sum_{j=1}^d a_{ij}}\right\rceil\right\}\right]-\mathbb{E}\left[1\wedge \exp\left\{\left\lceil\underset{i=1,...,n}{\max}\frac{\sum_{j=1}^d a_{ij}S_j-(k_i+m_i)}{\sum_{j=1}^d a_{ij}}\right\rceil\right\}\right],
\end{align*}

for $k_i>0,$ since for $k_i\leq 0,$ it is equal to $0$.
\\
\\
Finally, from \eqref{eq_AN_nleqslant_0} and defining $\boldsymbol{U}$ the random vector such that $U_i = \frac{\sum_{j=1}^d a_{ij} S_j -  m_i}{\sum_{j=1}^d a_{ij}}, \ \forall \ i = 1, \ldots, n,$ we get
\begin{equation*}
   \mathbb{P}(\boldsymbol{AN-m} \leq \boldsymbol{k}|\boldsymbol{AN} \nleqslant \boldsymbol{m})=
    1-\frac{\mathbb{E}\left[1\wedge \exp\left\{\left\lceil\underset{i=1,...,n}{\max}\left(U_i-\frac{k_i}{\sum_{j=1}^d a_{ij}}\right)\right\rceil\right\}\right]}{\mathbb{E}\left[1\wedge \exp\left\{\left\lceil\underset{i=1,...,n}{\max}\left(U_i\right)\right\rceil\right\}\right]}.
\end{equation*}
\end{proof}

\begin{proof}[\textbf{Proof of Proposition~\ref{prop:discrete_generalised_gpd}}]\label{pr:discrete_generalised_gpd}

Let $\boldsymbol\lceil\boldsymbol{Z}/\Lambda\rceil,$  we can write:
    \begin{align*}
    \mathbb{P}\left(\left\lceil\frac{\boldsymbol{Z}}{\Lambda}\right\rceil=\boldsymbol{k}\right)&=\mathbb{P}\left(\frac{\boldsymbol{Z}}{\Lambda}\leq \boldsymbol{k}\right)-\mathbb{P}\left(\frac{\boldsymbol{Z}}{\Lambda}\leq \boldsymbol{k-1}\right)
    =\mathbb{P}\left(\boldsymbol{Z}\leq \boldsymbol{k}\Lambda\right)-\mathbb{P}\left(\boldsymbol{Z}\leq (\boldsymbol{k-1})\Lambda\right)\\
&=\mathbb{E}\left(1\wedge e^{\max(\boldsymbol{S}-\alpha{log(\frac{\boldsymbol{k-1}}{\beta}+\boldsymbol{1})}}\right)-\mathbb{E}\left(1\wedge e^{\max(\boldsymbol{S}-\alpha log(\frac{\boldsymbol{k}}{\boldsymbol{\beta}}+\boldsymbol{1})}\right).
\end{align*}
Then, obviously, we get,
\begin{equation}\label{eq:ceil_Z_over_Lambda}
  \mathbb{P}\left(\left\lceil\frac{\boldsymbol{Z}}{\Lambda}\right\rceil\leq\boldsymbol{l}\right)=1-\mathbb{E}\left(1\wedge e^{\max(\boldsymbol{S}-\alpha log(\frac{\boldsymbol{l}}{\boldsymbol{\beta}}+\boldsymbol{1})}\right).  
\end{equation}

\noindent From Equation~\eqref{eq:ceil_Z_over_Lambda},  $\left\lceil\boldsymbol{Z}/\Lambda\right\rceil $ follows a $MDGPD(\boldsymbol{\frac{\beta}{\alpha}},\boldsymbol{\frac{1}{\alpha}},\boldsymbol{S}).$
\\
\\
Let $\boldsymbol{M}=\boldsymbol{Z}/\Lambda$, then we can write
\begin{align*}
    \boldsymbol{AM}\nleqslant \boldsymbol{m} 
    &\Leftrightarrow \exists i=1,\ldots,n,\, \sum_{j=1}^d a_{ij}M_j>m_i\\ 
    &\Leftrightarrow \exists i=1,\ldots,n, \, \sum_{j=1}^d a_{ij}N_j>m_i\Lambda\\
    &\Leftrightarrow  \exists i=1,\ldots,n,\, \sum_{j=1}^d a_{ij}E>m_i\Lambda - \sum_{j=1}^d a_{ij}S_j\\
    &\Leftrightarrow \exists i=1,\ldots,n, \,E>\frac{m_i}{\sum_{j=1}^d a_{ij}}\Lambda -\frac{\sum_{j=1}^d a_{ij}S_j}{\sum_{j=1}^d a_{ij}}.
\end{align*}

So, it comes 
\begin{align*}
    \mathbb{P}\left(\boldsymbol{AM}\nleqslant \boldsymbol{m}\right)&=\mathbb{E}\left(1\wedge\int_0^{+\infty}\exp\left(-\underset{1\leq i\leq n}{\min} \left(\frac{m_i}{\sum_{j=1}^d a_{ij}}x -\frac{\sum_{j=1}^d a_{ij}S_j}{\sum_{j=1}^d a_{ij}}\right)\right) \frac{\beta^{\alpha}}{\Gamma(\alpha)}e^{-\beta x}dx \right)\\
    &=\mathbb{E}\left(1\wedge\underset{1\leq i\leq n}{\max}\left(\exp\left(\frac{\sum_{j=1}^d a_{ij}S_j}{\sum_{j=1}^d a_{ij}}\right)\int_0^{+\infty}\frac{\beta^{\alpha}}{\Gamma(\alpha)}e^{-\left(\beta+\frac{m_i}{\sum_{j=1}^d a_{ij}}\right) x}\right)dx \right)\\
    &=\mathbb{E}\left(1\wedge\underset{1\leq i\leq n}{\max} \exp\left(\frac{\sum_{j=1}^d a_{ij}S_j}{\sum_{j=1}^d a_{ij}}-\alpha log\left(1+\frac{m_i}{\beta\sum_{j=1}^d a_{ij}}\right)\right)\right).
\end{align*}
Thus,
\begin{equation*}
    \mathbb{P}\left(\boldsymbol{AM}\leq \boldsymbol{m}\right)=1-\mathbb{E}\left(1\wedge\underset{1\leq i\leq n}{\max} \exp\left(\frac{\sum_{j=1}^d a_{ij}S_j}{\sum_{j=1}^d a_{ij}}-\alpha log\left(1+\frac{m_i}{\beta\sum_{j=1}^d a_{ij}}\right)\right)\right),
\end{equation*}
and
\begin{equation*}
    \mathbb{P}\left(\left\lceil\boldsymbol{AM}\right\rceil\leq \boldsymbol{m}\right)=1-\mathbb{E}\left(1\wedge\underset{1\leq i\leq n}{\max} \exp\left(\frac{\sum_{j=1}^d a_{ij}S_j}{\sum_{j=1}^d a_{ij}}-\alpha log\left(1+\frac{m_i}{\beta\sum_{j=1}^d a_{ij}}\right)\right)\right).
\end{equation*}
Finally, from Equation~\eqref{eq_AN_nleqslant_0}, we obtain
\begin{equation*}
    \mathbb{P}\left(\left\lceil\boldsymbol{AM}\right\rceil-\boldsymbol{m}\leq \boldsymbol{k}|\lceil\boldsymbol{AM}\rceil \nleqslant \boldsymbol{m}\right)= 1 - \frac{\mathbb{E}\left(1 \wedge \exp\left(\underset{i=1,\ldots,m}{\max}\frac{\sum_{j=1}^d a_{ij}S_j}{\sum_{j=1}^d a_{ij}} - \alpha \log\left(1 + \frac{k_i+m_i}{\beta\sum_{j=1}^d a_{ij}}\right)\right)\right)}{\mathbb{E}\left(1 \wedge \exp\left(\underset{i=1,\ldots,m}{\max}\frac{\sum_{j=1}^d a_{ij}S_j}{\sum_{j=1}^d a_{ij}} - \alpha \log\left(1 + \frac{m_i}{\beta\sum_{j=1}^d a_{ij}}\right)\right)\right)}.
\end{equation*}
Denoting by $\boldsymbol{U}$ the random vector with components $U_i = \frac{\sum_{j=1}^d a_{ij}S_j}{\sum_{j=1}^d a_{ij}} - \alpha \log\left(1 + \frac{m_i}{\beta\sum_{j=1}^d a_{ij}}\right)$,\\ $i = 1, \ldots,n$,
\begin{equation}
1 - \frac{\mathbb{E}\left[1 \wedge \exp\left\{\underset{i=1,\ldots,m}{\max} \left(U_i - \alpha \log\left(1 + \frac{k_i}{\beta \sum_{j=1}^d a_{ij}+m_i}\right)\right)\right\}\right]}{\mathbb{E}\left[1 \wedge \exp\left(\underset{i=1,\ldots,m}{\max} U_i \right)\right]}.
\end{equation}
\end{proof}

\begin{proof}[\textbf{Proof of Theorem~\ref{th:DS_over_cont}}]\label{pr:DS_over_cont}
Let $\boldsymbol{\sigma}>\boldsymbol{0}$ and $\boldsymbol{\xi}\geq\boldsymbol{0}$ be fixed for now.

For the discrete MDGPD as given in Theorem~\ref{def:non-standard_MDGPD}, where we denote the discrete spectral random vector by $\boldsymbol{S_D}$ (to avoid any confusion between the discrete and continuous version), the mass function at $\boldsymbol{k}$ is defined exactly by:
\[
p(\boldsymbol{k}) = \mathbb{E}\left(1 \wedge e^{\max\left(\boldsymbol{S_D} - \frac{\boldsymbol{1}}{\boldsymbol{\xi}} \log\left(\frac{\boldsymbol{\xi} (\boldsymbol{k-1})}{\boldsymbol{\sigma}} + \boldsymbol{1}\right)\right)}\right) - \mathbb{E}\left(1 \wedge e^{\max\left(\boldsymbol{S_D} - \frac{\boldsymbol{1}}{\boldsymbol{\xi}} \log\left(\frac{\boldsymbol{\xi} \boldsymbol{k}}{\boldsymbol{\sigma}} + \boldsymbol{1}\right)\right)}\right).
\]
Now let us compare the MDGPD and the discrete approximation of a continuous MGPD.
\\
Consider a continuous random vector $\boldsymbol{Y}\sim MGPD$ with cdf $F()$ defined by:
\[
F(\boldsymbol{x}) = 1 - \mathbb{E}\left(1 \wedge e^{\max\left(\boldsymbol{S} - \frac{\boldsymbol{1}}{\boldsymbol{\xi}} \log\left(\frac{\boldsymbol{\xi} \boldsymbol{x}}{\boldsymbol{\sigma}} + \boldsymbol{1}\right)\right)}\right), \text{ for $\boldsymbol{x}\in \mathbb{R}^d$},
\]
with $\boldsymbol{S}$ the continuous spectral random vector.

Then, the density of $\lceil\boldsymbol{Y}\rceil$ at $\boldsymbol{k} \in \mathbb{N}^{*d}$ can be approximated (see relation~\eqref{eq:approx_discrete_pdf}) by:
\[
f(\boldsymbol{k}) \approx F(\boldsymbol{k-1}) - F(\boldsymbol{k}),
\]
where $\boldsymbol{1}$ is the vector of ones as stated in the introduction.

Consider the ratio appearing in the Theorem~\ref{th:DS_over_cont}, say,
\[
R(\boldsymbol{k}, \boldsymbol{\sigma}) := \frac{\mathbb{E}\left(1 \wedge e^{\max\left(\boldsymbol{S} - \frac{\boldsymbol{1}}{\boldsymbol{\xi}} \log\left(\frac{\boldsymbol{\xi} \boldsymbol{k}}{\boldsymbol{\sigma}} + 1\right)\right)}\right)}{\mathbb{E}\left(1 \wedge e^{\max\left(\boldsymbol{S_D} - \frac{1}{\boldsymbol{\xi}} \log\left(\frac{\boldsymbol{\xi} \boldsymbol{k}}{\boldsymbol{\sigma}} + 1\right)\right)}\right)}.
\]
Since $\displaystyle \frac{\boldsymbol{\xi k}}{\boldsymbol{\sigma}} \underset{\boldsymbol{\sigma} \to \infty}{\to} \boldsymbol{0}$ uniformly over $k \in \mathbb{N}^*$, using that $\displaystyle \log(1 + x) = x + o(x)$ for $x$ close to $0$, 
we have, componentwise,
\[
\log\left(\boldsymbol{1 + \frac{\xi k}{\sigma}}\right) = \frac{\boldsymbol{\xi k}}{\boldsymbol{\sigma}} + o\left(\boldsymbol{\frac{1}{\sigma}}\right).
\]
Therefore, the arguments inside the exponential behave as:
\begin{equation}\label{eq:log_sigma_relation}
  -\boldsymbol{\frac{1}{\xi}} \log\left(\boldsymbol{1 + \frac{\xi k}{\sigma}}\right) \approx -\frac{\boldsymbol{k}}{\boldsymbol{\sigma}} + o\left(\frac{\boldsymbol{1}}{\boldsymbol{\sigma}}\right). 
\end{equation}
%

From the definitions of the standard MGPD and its discrete counterpart (Definition~\ref{def: st MGPD} and Theorem~\ref{th:MDGPD}), we know that both spectral vectors satisfy $\max(\boldsymbol{S})=\max(\boldsymbol{S_D})=\boldsymbol{0}$ from which we deduce that $$\displaystyle  e^{\max\left(\boldsymbol{S} - \frac{\boldsymbol{1}}{\boldsymbol{\xi}} \log\left(\frac{\boldsymbol{\xi} \boldsymbol{k}}{\boldsymbol{\sigma}} + \boldsymbol{1}\right)\right)}/ e^{\max\left(\boldsymbol{S_D} - \frac{\boldsymbol{1}}{\boldsymbol{\xi}} \log\left(\frac{\boldsymbol{\xi} \boldsymbol{k}}{\boldsymbol{\sigma}} + \boldsymbol{1}\right)\right)}\text{ goes to $\boldsymbol{1}$ as }  \boldsymbol{\sigma}\to \infty.$$

Moreover, since the function $x \mapsto 1 \wedge e^x$ is Lipschitz continuous with constant 1, the convergence is preserved under expectation:
\[
\left| \mathbb{E}\left[1 \wedge \exp\left(\max\left(\boldsymbol{S} - \frac{\boldsymbol{k}}{\boldsymbol{\sigma}}\right)\right)\right] 
- \mathbb{E}\left[1 \wedge \exp\left(\max\left(\boldsymbol{S_D} - \frac{\boldsymbol{k}}{\boldsymbol{\sigma}}\right)\right)\right] \right| \to 0,
\]
uniformly over $\boldsymbol{k}$ as $\boldsymbol{\sigma} \to \infty$.

Finally,
\[
\sup_{\boldsymbol{k} \in \mathbb{N}^{*d}} \left| R(\boldsymbol{k}, \boldsymbol{\sigma}) - 1 \right| \to 0, \quad \text{as } \boldsymbol{\sigma} \to \infty.
\]
\end{proof}

\subsection{Proof of Algorithm~\ref{algo:unknown_increments}}

In this section, we rely on the notation used in the bootstrap asymptotic theory by \cite{Bickel1981}.
Using the same notations as in Algorithm~\ref{algo:known_increments}, we first consider a sample $(\Delta_1,\ldots,\Delta_n)$ drawn from a random variable $\Delta$ with cdf $F$ and with empirical estimation cdf $F_n.$ From $(\Delta_1,\ldots,\Delta_n),$ we simulate $\Delta^*_1,\ldots,\Delta^*_m.$ Let $F_n^{(m)}$ the empirical cdf of $(\Delta^*_1,\ldots,\Delta^*_m)$ representing a sample of the random variable $\Delta^*.$

Let us prove the following lemma that guarantees the convergence of Algorithm \ref{algo:unknown_increments}. 
\begin{lemma}
If $F_n^{(m)}$ converges in distribution to $F$, as $n$ and $m$ tend to infinity, then $(N^*_{1,k},N^*_{2,k} )_{1\leq k \leq m}$ converges in distribution to a bivariate $DGPD$ $\boldsymbol{N}$ from which is drawn the original sample $(N_{1,i},N_{2,i})_{1\leq i\leq n}.$
\end{lemma}

\begin{proof}
We know that $\mathbb{P}(\max(\boldsymbol{N})\leq q)=1-1\wedge e^{-q},$ for any $q\in \mathbb{N}^*.$ The distribution of $(N^*_1,N^*_2 )$ satisfies
\begin{align*}
    \mathbb{P}\left(N^*_{1}\leq n_1, N^*_{2}\leq n_2\right) &= \mathbb{P}\left(\max(\boldsymbol{N})\leq n_1-\Delta^*\mathds{1}_{\left(\Delta^*<0\right)},\, \max(\boldsymbol{N})\leq n_2+\Delta^*\mathds{1}_{\left(\Delta^*\geq 0\right)}\right)\\
    &= 1 - \mathbb{P}\left(\max(\boldsymbol{N}) > \min (n_1-\Delta^*\mathds{1}_{\left(\Delta^*<0\right)},n_2+\Delta^*\mathds{1}_{\left(\Delta^*\geq 0\right)}\right) \\
    &=1-\mathbb{E}\left(1\wedge e^{-\min\left(n_1-\Delta^*\mathds{1}_{\left(\Delta^*<0\right)},n_2+\Delta^*\mathds{1}_{\left(\Delta^*\geq 0\right)}\right)}\right)\\
    &=1 - \mathbb{E}\left(1 \wedge e^{-\min(n_1-\Delta^*, n_2) - \max(0, \Delta^*)}\right).
\end{align*}
\noindent Then, as $n,m\rightarrow \infty$,

\begin{equation*}\label{eq:T-max(T)_MDGPD}
\begin{aligned}
\mathbb{P}\left(N^*_{1}\leq n_1, N^*_{2}\leq n_2\lvert (\Delta_1,\ldots,\Delta_n) \right) &\rightarrow 1 - \mathbb{E}\left(1 \wedge e^{-\min(n_1-\Delta, n_2) - \max(0, \Delta)}\right) \\
&=1-\mathbb{E}\left(1\wedge e^{\max(T_1-n_1,T_2-n_2)-\max(T_1,T_2)}\right),
\end{aligned}
\end{equation*}

which is the cdf of the bivariate $DGPD$.
\end{proof}

\subsection{Some extension to dimension $3$}\label{app:dim_3}

If $d=3,$ we write $\boldsymbol{N}=(N_1,N_2,N_3)$ and $\boldsymbol{T}=(T_1,T_2,T_3).$ Define $\Delta_i=T_i-\max\{T_j:j\neq i\}$ for $i=1,2,3.$ Thus, we can write 
\begin{equation*}
\begin{cases}
    N_1 = \max(\boldsymbol{N}) + \Delta_1 \mathds{1}_{(\Delta_1 < 0)}, \\
    N_2 = \max(\boldsymbol{N}) + \Delta_2 \mathds{1}_{(\Delta_2 < 0)}, \\
    N_3 = \max(\boldsymbol{N}) + \Delta_3 \mathds{1}_{(\Delta_3 < 0)}.
\end{cases}
\label{eq:combined}
\end{equation*}

\begin{algorithm}
\caption{Bootstrap $MDGPD$ simulation for $d=3$}
\label{algo:unknown_increments_d=3}
\begin{algorithmic}[1]
\Require A sample of $(N_{1,i},N_{2,i},N_{3,i})_{1\leq i\leq n}\sim MDGPD$ as described
\Ensure A simulated sample $(N^*_{1,k},N^*_{2,k},N^*_{3,k})_{1\leq k\leq m}$

\State Define $\Delta_{l,i} = T_{l,i} - \max\{T_{l,j}: j \neq i\}$ for $l=1,2,3$
\State Generate $m$ realizations $\max(\boldsymbol{N})_k\sim \text{Geom}(1-e^{-1})$ independently from $\Delta_i$
\State Bootstrap $m$ realizations $(\Delta^*_{1,k},\Delta^*_{2,k},\Delta^*_{3,k})$ from $(\Delta_{1,i}, \Delta_{2,i}, \Delta_{3,i})_{1\leq i \leq n}$
\State \textbf{Return} $N^*_{1,k} = \max(\boldsymbol{N})_k + \Delta^*_{1,k}\mathds{1}_{\left(\Delta^*_{{1,k} < 0}\right)},$
\Statex \hspace{1.7cm} $N^*_{2,k} = \max(\boldsymbol{N})_k + \Delta^*_{2,k}\mathds{1}_{\left(\Delta^*_{{2,k} < 0}\right)},$
\Statex \hspace{1.7cm} $N^*_{3,k} = \max(\boldsymbol{N})_k + \Delta^*_{3,k}\mathds{1}_{\left(\Delta^*_{{3,k} < 0}\right)},$
\Statex \hspace{1.7cm} for $1\leq k \leq m.$
\end{algorithmic}
\end{algorithm}

\noindent In the same way as in dimension $2,$ we can provide an illustration using simulation from a parametric model with trivariate independent Poisson random variables as generator $\boldsymbol{T}$.

\begin{figure}[htbp]
    \centering
    \begin{subfigure}[b]{0.48\textwidth}
        \centering
        \includegraphics[width=\textwidth]{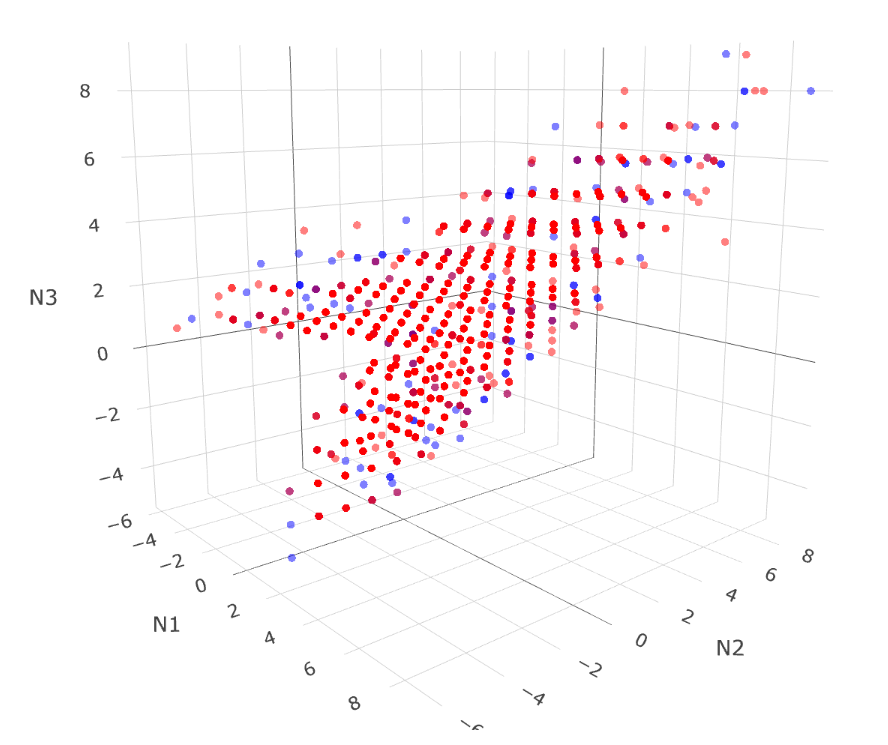}
        \caption{Simulated trivariate data}
        \label{fig:modeld3_image1}
    \end{subfigure}
    \hfill
    \begin{subfigure}[b]{0.51\textwidth}
        \centering
        \includegraphics[width=\textwidth]{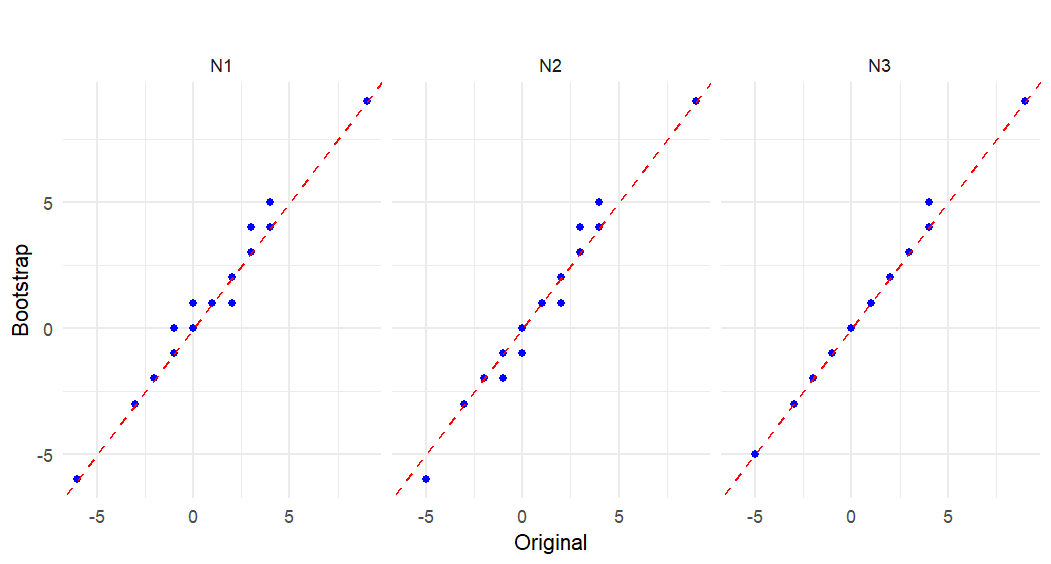}
        \caption{Quantile-Quantile plots}
        \label{fig:modeld3_image2}
    \end{subfigure}
    \caption{\sf From left to right : (1) Scatter plot of simulated data of trivariate MDGPD with Poisson ($1$) generator of sample size $n=10000$ (blue dots) and sampled data from one simulation using Algorithm \ref{algo:unknown_increments_d=3} with sample size $m = 10000$ (red circles) (2) Quantile-Quantile plots for $N_1,$ $N_2$ and $N_3$.}
\end{figure}   

Figure~\ref{fig:modeld3_image2} displays a good adequation between the original simulated marginals and bootstrap simulations.



\end{document}